\documentclass[11pt,preprint]{aastex}

\begin{document}

\newcommand\etal{et al. }

\def\t0{\theta_{\circ}}
\def\muo{\mu_{\circ}}
\def\sd{\partial}
\def\be{\begin{equation}}
\def\en{\end{equation}}
\def\bv{\bf v}
\def\bvo{\bf v_{\circ}}
\def\ro{r_{\circ}}
\def\rhoo{\rho_{\circ}}
\def\etal{et al.\ }
\def\msun{M_{\sun}}
\def\rsun{R_{\sun}}
\def\lsun{L_{\sun}}
\def\msunyr{M_{\sun} \, yr^{-1}}
\def\kms{\rm \, km \, s^{-1}}
\def\mdot{\dot{M}}
\def\ha{H$\alpha \;$}
\def\ecs{\rm erg \, cm^{-2} \, s^{-1}}

\title{The CIDA Variability Survey of Orion OB1. I: 
the low-mass population of Ori OB 1a and 1b\altaffilmark{1,2}}
\author{Cesar Brice\~no\altaffilmark{3,7}, 
Nuria Calvet \altaffilmark{4},
J. Hern\'andez\altaffilmark{3,5},
A.K. Vivas\altaffilmark{3}, 
Lee Hartmann\altaffilmark{4}, 
J.J. Downes\altaffilmark{3,6},
and
Perry Berlind\altaffilmark{4}}

\altaffiltext{1}{Based on observations obtained at the Llano del Hato
National Astronomical Observatory of Venezuela, operated by CIDA for the
Ministerio de Ciencia y Tecnolog{\'\i}a, and at the Fred Lawrence Whipple
Observatory (FLWO) of the Smithsonian Institution.}
\altaffiltext{2}{Based on observations obtained at the 3.5m WIYN telescope.
The WIYN Observatory is a joint facility of the University of Wisconsin-Madison, Indiana University, Yale University, and the National Optical Astronomy Observatory.}
\altaffiltext{3}{Centro de Investigaciones de Astronom{\'\i}a (CIDA),
    Apartado Postal 264, M\'erida 5101-A, Venezuela;
Electronic mail: briceno@cida.ve,avivas@cida.ve,jesush@cida.ve}

\altaffiltext{4}{Smithsonian Astrophysical Observatory, Mail Stop 42, Cambridge, MA 02138, USA;
Electronic mail: ncalvet@cfa.harvard.edu,hartmann@cfa.harvard.edu,pberlin@cfa.harvard.edu}

\altaffiltext{5}{Also at the Postgrado de F{\'\i}sica Fundamental,
Universidad de los Andes, M\'erida, Venezuela}

\altaffiltext{6}{Also at the Postgrado en F{\'\i}sica, Universidad Central de Venezuela, Caracas,  Venezuela}

\altaffiltext{7}{Visiting Astronomer, Kitt Peak National Observatory, National Optical Astronomy Observatory, which is operated by the Association of Universities for Research in Astronomy, Inc. (AURA) under cooperative agreement with the National Science Foundation. }

\begin{abstract}
We present results of a large scale, multi-epoch optical
survey of the Ori OB1 association, carried out with the QuEST camera
at the Venezuela National Astronomical Observatory.
We identify for the first time the widely spread low-mass,
young population in the Orion OB1a and OB1b sub-associations.
Candidate members were picked up by their
variability in the $V$-band and position in 
color-magnitude diagrams. We obtained spectra to confirm membership.
In a region spanning $\rm \sim 68 \Box^\circ$
we found 197 new young stars; 
of these, 56 are located in the Ori OB1a subassociation and 
142 in Ori OB1b.
The spatial distribution of the the low mass young stars
is spatially coincident
with that of the high mass members,
but suggests a much sharper edge to the association.
Comparison with the spatial extent of
molecular gas and extinction maps indicates
that the subassociation Ori 1b is concentrated 
within a ring-like structure of radius $\sim$ 2$^\circ (\sim$
15 pc at 440 pc), centered roughly
on the star $\epsilon$ Ori in the Orion belt. 
The ring is apparent in $^{13}$CO and corresponds to
a region with an extinction $A_V \ge 1$.
The stars exhibiting strong H$\alpha$ emission, an
indicator of active accretion,
are found along this ring,
while the center is populated with weak H$\alpha$ emitting stars.
In contrast, Ori OB1a is located
in a region devoid of gas and dust.
We identify a grouping of stars within a $\rm \sim 3\> deg^2$ 
area located in 1a, roughly clustered around the B2 star 25 Ori. 
The Herbig Ae/Be star V346 Ori is also associated 
with this grouping,
which could be an older analog of $\sigma$ Ori.
Using using several sets of evolutionary
tracks we find an age of 7 - 10 Myr for
Ori 1a and of $\sim 4 - 6$ Myr for Ori OB1b,
consistent with previous estimates from OB stars.
Indicators such as the equivalent
width of H$\alpha$ and near-IR excesses
show that the number of accreting low-mass stars
decreases sharply between  Ori 1b and Ori 1a.
These results indicate that while a substantial
fraction of accreting disks remain at ages $\sim 5$ Myr,
inner disks are essentially dissipated by 10 Myr.
\end{abstract}

\keywords{ stars: formation ---  stars: pre-main sequence
--- surveys}

\section{Introduction}

The Orion OB1 Association, located at roughly 400 pc (see review by
\citealt{ges89})
and spanning over $\rm 200 \> deg^2$ on the sky, 
is one of the largest and nearest regions with active star formation.
With a wide range of ages and environmental conditions, Orion
is an ideal laboratory for investigating fundamental questions
related to the birth of stars and planetary systems.
\citet{bla64} identified four major subgroups
which differ in age and their content of gas and dust: Ori OB1a, 1b (that roughly
encompasses the belt stars $\delta$, $\epsilon$ and $\zeta$ Ori), 1c and 1d
(where the Orion Nebula Cluster [ONC] is located).
Within this vast region there are extremely young groupings of stars still embedded in
their parent clouds, such as NGC 2068, 2071, 2023 and 2024 in the Orion B
molecular cloud, observed mostly at IR and sub-mm wavelengths (\citealt{dal99}; 
\citealt{hlp01}; \citealt{maw01}); 
the young, dense, and partially embedded ONC
\citep[hereafter H97]{het86,ali95,hil97}; 
clusterings of intermediate age ($\sim 3-5$ Myr) stars including
the $\sigma$ Ori \citep{gar67,wwf97} 
and $\lambda$ Ori groups \citep{mup77,dom99,dom01,dom02}
and older populations like the $\sim 10$ Myr old OB1a sub-association, 
located in an area essentially devoid of gas and dust.

While the earliest stages of stellar evolution
must be probed with infrared and radio
techniques, many fundamental questions including lifetimes 
of molecular clouds, cluster dispersal, protoplanetary disk
evolution and triggered star formation, can only
be addressed by optical surveys of older populations with
ages $\sim$ 3-20 Myr.
To make progress on these issues, wide-field studies of stellar populations
in and near star-forming regions are required.  In the past, studies of
OB associations have been used to investigate sequential star formation
and triggering on large scales (e.g., \citealt{bla91} and references therein).
However, OB stars are formed essentially on the main sequence
(e.g. \citealt{pas92,pas93}) and evolve off the main sequence on 
timescales of order 10 Myr (depending upon mass
and amount of convective overshoot); thus, they are not useful tracers of
star-forming histories on timescales of 1-3 Myr.
Moreover, we cannot investigate cluster structure and dispersal or
disk evolution without studying low-mass stars.  
Wide-field studies of the low mass populations
have been carried for some portions of the Upper
Scorpius OB association \citep{prz99,pbb02}. 
In Orion, many young
individual clusters have been studied at both optical and infrared
wavelengths (e.g. \citealt{lad92,psh94,hil97,smm99}), 
but these only represent the highest-density, and often youngest, regions.
Little is known about the more widely spread low mass stellar population in
these OB associations.

A few large scale surveys have tried to map the widely distributed
low-mass pre-main sequence population in 
Orion.
The Kiso objective-prism H$\alpha$ survey \citep{wky89,wky91,wky93,kyw89,nwk95}, 
covered 225$\Box^\circ$ around the clouds
but was biased towards the strongest H$\alpha$ emitting stars.
The ROSAT All-Sky Survey \citep{atw96}
detected many X-ray strong young stars widely distributed across Orion; 
however, this selection was strongly affected by
older foreground field stars unrelated to the young population of Orion
\citep{bhs97} and because of its shallow limit it did
not reach the fainter K and M stars.
\citep{wws98} carried out a detailed
photometric and spectroscopic study of the region
around the O9.5 type star $\sigma$ Ori, belonging to 1b, and found a
large concentration of pre-main sequence
low mass stars down to V $\sim$ 19.
Similar high densities of young low mass stars
have been found in other selected small regions
in 1a and 1b by \citet{sww99} and \citet{she03},
indicating that many low mass stars 
remain undiscovered.

Until recently, it was not possible to conduct
unbiased, large-scale optical surveys
but technological advances have now made it possible,
building on the availability of cameras with multiple CCDs
and multiobject spectrographs on telescopes with wide fields of view.
One such survey is the optical variability study we have conducted over
$\rm \sim 173\> deg^2$ in the Orion OB1 association
with the purpose of  finding, mapping, and studying large numbers of widely-dispersed,
low mass ($\la 1 \msun$) stars with ages from 1 to $\sim$ 20 Myr.

Optical photometric variability is one of the defining characteristics
of pre-main sequence stars \citep{joy45,her62}. However, due to
limitations in the size of CCD detectors, past work on
photometric variability of young stars have concentrated on follow up
studies of selected samples, that had been identified as pre-main
sequence objects by some other means. Only very recently has the
potential of variability as a technique to pick out young stars
amongst the field population started to be realized \citep{bvc01,lbm04}.

In this contribution 
we present results from our variability survey in 
an area of nearly $\rm 68\> deg^2$, that
includes the Ori OB1b sub-association, a large fraction of Ori OB1a,
and part of the Orion B molecular cloud \citep{mmm87}, with
the young clusters NGC 2023, 2024, 2068 and 2071.
In \S 2 we describe the optical variability survey and follow up
spectroscopy. In \S 3 we describe our results,
and in \S 4 we present a discussion and conclusions.

\section{Observations and Data Reduction}
\label{sec_obs}

\subsection{The QuEST Camera}
\label{sec_quest}

The multi-band, multi-epoch photometric survey was carried out using an
$8000 \times 8000$  CCD Mosaic Camera built by the QuEST
collaboration{\footnotemark[1]}\footnotetext[1]{ The QuEST (Quasar
Equatorial Survey Team) collaboration \citep{sny98} included Yale
University, Indiana University,
Centro de Investigaciones de Astronom{\'\i}a, and Universidad de Los
Andes (Venezuela). The main goal of QuEST was to perform a large scale
survey of quasars, but the impressive capabilities of this instrument
prompted a variety of other studies.}
 and installed on the 1m (clear aperture)
Schmidt telescope at the Llano del Hato National Astronomical Observatory,
located at an elevation of 3610 m in the Venezuelan Andes,
close to the equator ($8^\circ 47'$ N) and under dark skies.
The 16 $2048\times 2048$ Loral CCD devices are set in a 4x4 array
covering most of the focal plane of the Schmidt telescope. The chips
are front illuminated with a pixel size of $15\mu$m, which 
corresponds to a scale of 1.02'' per pixel. Because of the
gaps between rows of CCDs in the east-west direction, the
effective field of view is $\rm 5.4 deg^2$
(see \citealt{bsa02} for a detailed description of the instrument).

The camera has been designed for observing in drift-scan
mode: the telescope is fixed and the CCDs are read out E-W at the
sidereal rate as stars drift across the device.
This procedure results in the generation of a continuous
strip (or ``scan'') of the sky, $2.3^\circ$ wide; 
the sky can be surveyed at a rate of $34.5\Box^o/hr$.
In addition to its high efficiency for covering large areas
in a short time, drift-scanning offers other significant advantages:
the telescope does not have to be moved so there is no time lost for
reading out the CCDs. Also, flat-fielding errors are minimized, since each
stellar image is obtained by averaging over many pixels as it travels
over the CCD,
and astrometric precision is enhanced because the zenith angle is
constant throughout.
One of the limitations of drift-scanning is that the sky motion is only
parallel on the equator and the rate of drift on the focal plane
is a function of the declination. To address this
effect, each row of CCDs is mounted on a separate ''finger'' which
pivots at one end, allowing the row to be independently aligned
perpendicular to the star paths by moving an eccentric cam at
the opposite end.  The chips are clocked out
at separate rates to compensate for the change in sidereal rate
as a function of declination;
even so, the size of the devices and the image scale limit the
declination range available for drift-scanning. To minimize
image smearing to $\le 0.5$ pixel, drift-scanning is restricted
to $-6^\circ \le \delta \le +6^\circ$.
The drift-scan mode allows to obtain
quasi-simultaneous multi-band photometry since each column of chips,
oriented N-S, can be fitted with separate filters;
stars will successively cross over 4 chips in a column each with its own filter{\footnotemark[2]}
\footnotetext[2]{In August 2001 a lightning struck the 1m Schmidt dome and damaged the
CCD Mosaic Camera, resulting in the loss of an entire row of CCDs. Therefore,
starting in the Fall of 2001 we only could obtain data simultaneously in 3 filters.}.
Several filters are available, including BVRI and $\rm H\alpha$.
The exposure time (hence the limiting magnitude) are
set by the time a star needs to cross a single chip (140 sec at
$\delta = 0^\circ$). Deeper images can be obtained by combining
several scans. 
This combination of exposure time with the typical seeing at the site
(see below) result in a
$10\sigma$ limiting magnitude of $\sim 19.5$ in the V-band.
Saturation occurs at $V = 13.0 - 14.5$ depending on the CCD,
since they have differing well depths.
The data acquisition system is described in Sabbey et al. (1998).

\subsection{The Optical Variability Survey}

The location of the telescope, together with the advantages of
the drift-scan technique make this an ideal survey instrument
for large scale studies of regions close to the celestial
equator like Orion.
Our photometric survey spans the region from
5h - 6h in RA and $\rm DEC= -6^\circ to +6^\circ$, shown in Figure 1 of \citet{bvc01}.
We covered this large portion of the sky in 6 strips at constant declination,
centered at $\rm DEC= -5^\circ, -3^\circ, -1^\circ, +1^\circ, +3^\circ 
and +5^\circ$, for a total area of $\rm \sim 173\> deg^2$ (because of the 
$\sim 50\arcsec$ gaps between CCDs in DEC, we lose about $\rm 1.1\> deg^2$ per scan).
Here we present results for two strips at $\rm DEC= -1^\circ \> and \> +1^\circ$,
encompassing an area of nearly $\rm 68\> deg^2$, and a spanning the range 
$l \rm \sim 200^\circ \> to \> \sim 210^\circ$ and 
$\rm \vert b \vert = -10^\circ \> to \> -22^\circ$.
This part of Orion
includes the region around the belt stars $\delta$, $\epsilon$, and
$\zeta$, that constitutes the OB1b sub-association \citet[hereafter WH]{wah77,wah78}
the region north and west of it, the OB1a sub-association, and 
part of the Orion B molecular cloud \citep{mmm87}, with
the young clusters NGC 2023, 2024, 2068 and 2071.

Our scans start on average a little before $\rm RA=5^h$ and
end a little beyond $\rm RA= 6^h$, in order to have complete coverage
in all filters in the range $\rm RA=5^h  - 6^h$.
Some of our observations were shared with other
projects, therefore a number of our scans, 
especially at $\rm DEC= -1^\circ$, start at RA $\rm \sim 3-4^h$.
In practice, the start time and length of each drift scan was determined by the
sky conditions; some scans had to be cut short of $\rm RA= \sim 6^h$.
The temporal spacing between observations was mostly determined by weather
and scheduling constraints because of other ongoing projects with the instrument.
We obtained 23 scans at $\rm DEC= -1^\circ$ during 18 nights
from December 1998 through January and February 1999 and then from 
late 1999 through early 2000;
the 33 scans at $\rm DEC= +1^\circ$, collected over 17 nights, are
more sparsely distributed in time,
starting in Dec. 1999, continuing in October 2002, January and February 2003,
with the final data obtained in October 2003.  
The observation log is shown in Table \ref{tabcida}.
The general observational strategy was to observe as close to the meridian
as possible with a maximum hour angle of $\rm 2\fh 5$; 
because of this and the equatorial location of Llano del Hato,
the airmass for most of our observations is $\la 1.3$. 
The typical seeing during our observations ranged
from $1.8\arcsec$ to $\sim 2.5\arcsec$; however, due to the expected
degradation introduced by drift-scanning \citep{gih92}
combined with the pixel size, most scans have images with FWHM $\sim 2.5-3\arcsec$.
Thus, with a pixel scale of $1\farcs 02$ the point spread
function is well sampled.

\subsection {Data Processing and Variable Star Identification}

In order to effectively process the approximately 30 Gb/night of raw data produced
by this instrument, QuEST developed its own software.
The whole process is completely automated with minimum interaction from
the user.  The QuEST offline software performs the
basic reduction: overscan subtraction and flat-fielding (one-dimensional flats
are used to correct the response variations along the N-S direction);
object detection and seeing determination (by measuring an average
FWHM for $\sim 10^2 - 10^3$ of stars in each frame); determines the positional 
offsets between CCDs (recognizes a single star in all 4 fingers - rows);
performs aperture photometry with an aperture radius based on the
FWHM of the particular frame (several apertures are used:
1$\times$FWHM, 2$\times$FWHM, etc.); finally, Astrometry is done
by matching the objects with the USN-A2.0 catalog \citep{mon98}.
Positions are internally accurate to $\pm 0.2\arcsec$.
Each catalog for every drift scan
is stored in binary format and contains, among others,
positions J2000.0, instrumental magnitudes in 4 bands,
errors, fluxes within several apertures, flags (bad columns, edges, etc.).
A detailed description of the data pipeline is given in \citet{rma04}.

We have used differential photometry to identify variable stars among
this large number of objects. 
Our main interest is picking out candidate pre-main sequence stars
among the general field population. Low-mass ($\rm M \la 2\> \msun$) 
young stars (T Tauri stars, hereafter TTS) are known to vary on time scales from
hours to weeks, with variations of up to a few magnitudes. 
Usually the variations are larger in the $V$-band than in the $I$-band.
This variability can be irregular,
thought to originate in flare-like activity or in non-steady accretion
for the strong H$\alpha$ emitting objects (Classical T Tauri stars, CTTS);
the periodic variations, seen mostly in weak H$\alpha$ emitting TTS (WTTS),
are proposed to be the result of rotational modulation by hot or cool 
stellar spots in the star's surface \citep{hhg94}
Therefore, our choice of the $V$ filter allows us to
sample a wavelength range were TTS are expected to vary the most.
Also, because we are not aiming at obtaining light curves or deriving 
periods for our stars, our irregular temporal sampling is not
a major limitation, in fact, since our data spans time scales
from hours (in a few cases), to days, months and years, 
we expect to detect most of the TTS within our survey area and 
down to our sensitivity limits.

In order to derive the relative photometry 
we first chose as reference catalogs those
corresponding to scans obtained during clear nights, which
showed a stable behavior of the sky background as a function of RA
as measured from the average value for the ensemble photometry of thousands
of stars in each frame.
For the DEC=$-1^\circ$ strip we used observation No. 512 from January 06, 1999,
and observation No. 502 from December 26, 1999 for DEC=$+1^\circ$.
After doing a positional matching, within a $2\arcsec$ search radius,
of all stars in each reference scan with respect to the
other scans, we obtained a final combined catalog for the 
DEC=$-1^\circ$ and $+1^\circ$ strips containing 340,532 objects,
which had a minimum of three good measurements. This master catalog
incorporated data from 21 out of the 23 scans at $\rm DEC=-1^\circ$ 
and 31 of the 33 scans at $\rm DEC=+1^\circ$. Some scans had to be
discarded because of excessive cloud cover during the observation.

The instrumental magnitudes were normalized to each reference
scan in order to remove variable atmospheric extinction due
to differences in air mass and sky transparency. This
normalization to reference magnitudes was done in bins of
$0.25^\circ$ of right ascension, equivalent to $\sim 1$ minute of time
in drift-scan mode. Then we calculated the magnitude differences 
between each scan and the reference scan for all stars in each bin
of right ascension, and an ensemble clipped mean was produced.
Since each bin typically contains several hundred stars, especially
at the relatively low galactic latitude of our survey area,
this procedure gives a robust estimate of the zero-point difference
between each scan and the reference one. With this approach, variable
sky conditions like a passing cloud causing an extinction of up to 
1 magnitude can be corrected, as shown in Figure 3 of
\citet{vza04}, where more details of this method are provided.

\begin{figure}
\plotone{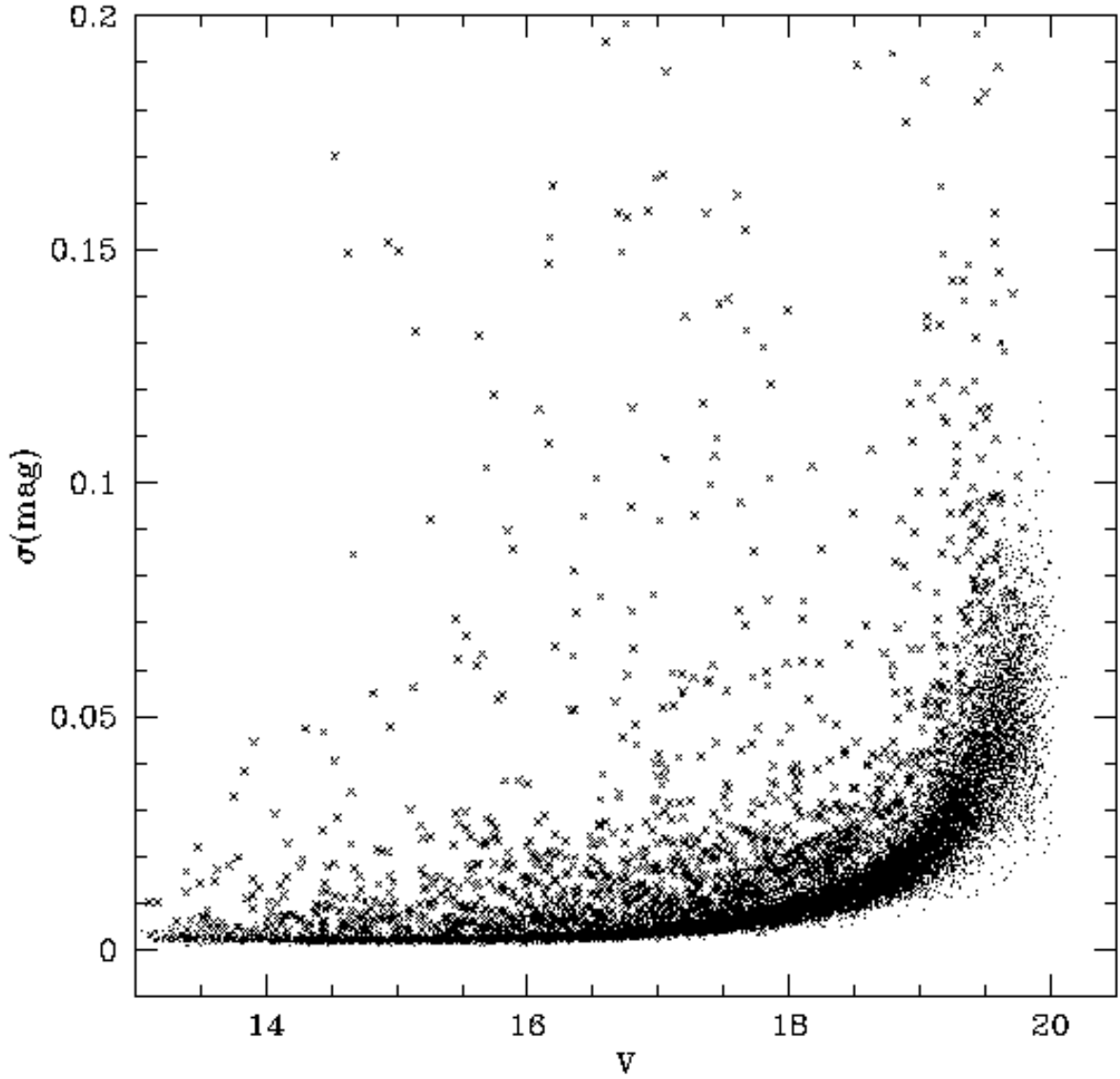}
\caption{The $1\sigma$ dispersion of up to 32 $V$-band magnitude
measurements for $\sim 30000$ stars in a $\rm 5.3 \> deg^2$ 
portion of the Orion variability survey. 
Crosses indicate the stars selected as variable by the $\chi^2$ test.
}
\label{fig_sig}
\end{figure}

Since a large number of stars have been measured, the random errors 
of the differential magnitudes are well determined. These amount to
$\sim 0.005$ magnitudes for $V \le 17$ and increase to $\sim 0.08$ mag
at $V\sim 19.7$ as photon statistics dominate the noise. 
Because of the larger errors when progressing toward 
the limiting magnitude of our observations,
fainter variables are biased toward larger values of $\rm \Delta V$.
In Figure \ref{fig_sig} we show the dispersion in the normalized 
instrumental $V$-band magnitudes for a subsample of the data 
corresponding to an area of $\rm 5.2\> deg^2$
containing $\sim 30,000$ stars.
Each point represents the statistics of up to 32 measurements.
Objects detected as variables are indicated with an "x" mark.
Most objects are non-variable and populate the curved region; 
points above the curve are likely variable stars.

Following \citet{vza04} we applied a $\chi^2$ test on the
normalized magnitudes to select variable stars, which is
appropriate because non-variable stars follow a Gaussian
distribution of errors \citep[Figure 5 in][]{vza04}
If the probability that the
dispersion is due to random errors
is very low ($\le 0.01$), the object
is flagged as variable.
Before running the $\chi^2$ test we eliminated measurements
that are potentially affected by  cosmic rays, by deleting
points that were more than $4\sigma$ away from the mean
magnitude for that particular object.
We flagged all objects with a 99\% confidence level of
being variable [$\rm P(\chi^2) \le 0.01$]. 
This confidence level
sets the minimum $\Delta {\rm mag}$ that we can detect as a function of
magnitude: 0.06 for $V=15$, 0.09 for $V=17$ and
0.30 for $V=19$.
This means that for fainter stars we detect only those which vary the most.
We cannot decrease the confidence level very much without increasing the
contamination from non-variable stars to unacceptable levels because we are
studying hundreds of thousands of stars.
Following this scheme
we detected 20729 variables in the $\rm DEC=-1^\circ$ and
$\rm DEC=+1^\circ$ strips, a $\sim 6$\%  
fraction of variable stars. The selection criterion we
used still allows $\sim 3400$ possible spurious variables.
The fraction of variables we find is larger than that found
by other variability surveys, e.g. 0.4\% in the All Sky 
Automated Survey \citep{pojmanski03}, and 0.8\% in the ROTSE
variability survey \citep{akerlof00}.
However, the fraction of false variables in our data may be
quite higher than the number expected from the confidence level
of the $\chi^2$ test alone.
This could be the result of close neighbors and blends, 
especially at the galactic latitude of Orion.
We did not attempt to identify these false variables because
our selection criteria include additional constraints, particularly
spectroscopy, that will further reject contaminants.
At this point we proceeded to plot the variables in
color-magnitude diagrams to select pre-main sequence candidates.

For the calibration of the photometry we used the
Smithsonian Astrophysical Observatory (SAO) 1.2m telescope 
equipped with the 4SHOOTER CCD Mosaic Camera
\citep{caf93}
\footnote{http://linmax.sao.arizona.edu/help/FLWO/48/4CCD.primer.html}
at Mt. Hopkins, Arizona,
during the nights of Dec. 3 and 4, 2002.
The 4SHOOTER
is an array of four CCDs, covering a square of $\sim$25 arcminutes side, with the
four chips of approximately 12 arcminutes side. We used a binning of $2\times 2$
yielding a plate scale of 0.66 \arcsec/pix.
We obtained $V$ and Cousin $I$-band images of selected fields at RA=5h-6h, with
declinations such that they would fall on parts of the sky overlapping
each of the four columns in our QuEST Camera scans.
Throughout each night we observed the
\citet{landolt92} standard fields PG0231+051, SA 92 and SA 98 at airmasses. 1-1.5.
We centered all our target fields on chip 3, the detector with best cosmetics.
Raw images were bias-subtracted and flattened (with dome flats)
using standard IRAF\footnote{IRAF is distributed by the National Optical Astronomy
Observatories, which are operated by the Association of Universities for Research
in Astronomy, Inc., under cooperative agreement with the National Science Foundation} routines.
We then performed aperture photometry on all our frames. Typically we had 12-20
standard stars per night.
Our photometric solution was derived with a rms of 0.02 mags for $V$ 
and 0.03 mags for $I_C$, 
In each of the secondary standard fields
we selected stars in the range $V=15-18$, much fainter than our
saturation limit, and bright enough to yield a good SNR. 
Positions within 0.5\arcsec were obtained using the CM1 program \citet{stockabad88}
to produce a plate solution for each frame, after comparing with $\sim 6$ reference
stars from the Guide Star Catalog.\footnote{
The Guide Star Catalogue II is a joint project of the Space Telescope Science Institute 
and the Osservatorio Astronomico di Torino. Space Telescope Science Institute is operated 
by the Association of Universities for Research in Astronomy, for the National Aeronautics 
and Space Administration under contract NAS5-26555. The participation of the Osservatorio 
Astronomico di Torino is supported by the Italian Council for Research in Astronomy. 
Additional support is provided by  European Southern Observatory, Space Telescope European 
Coordinating Facility, the International GEMINI project and the European Space Agency 
Astrophysics Division.} 
This accuracy is enough for positional matching with our
QuEST Camera catalog.
Finally, we kept only those stars with more than 7 measurements and 
not flagged as variable by our variability selection
scheme (see below). Our final list contains 390 secondary standards,
a robust set of comparison stars which we used to calibrate the normalized
magnitudes computed by our variability programs from the QuEST Camera scans.

\subsection{Selection of candidate pre-main sequence stars}

Traditional multicolor photometric surveys for pre-main sequence stars 
have relied mainly on color-magnitude and color-color diagrams to select
candidate young stars. This approach has been feasible for
previous studies that dealt with regions spanning small
areas. However, 
because of the large number of potential
pre-main sequence stars in our survey, and the contamination
by large numbers of field stars, this selection
method is too inefficient.
We therefore use variability to
select candidate association members.

\begin{figure}
\plottwo{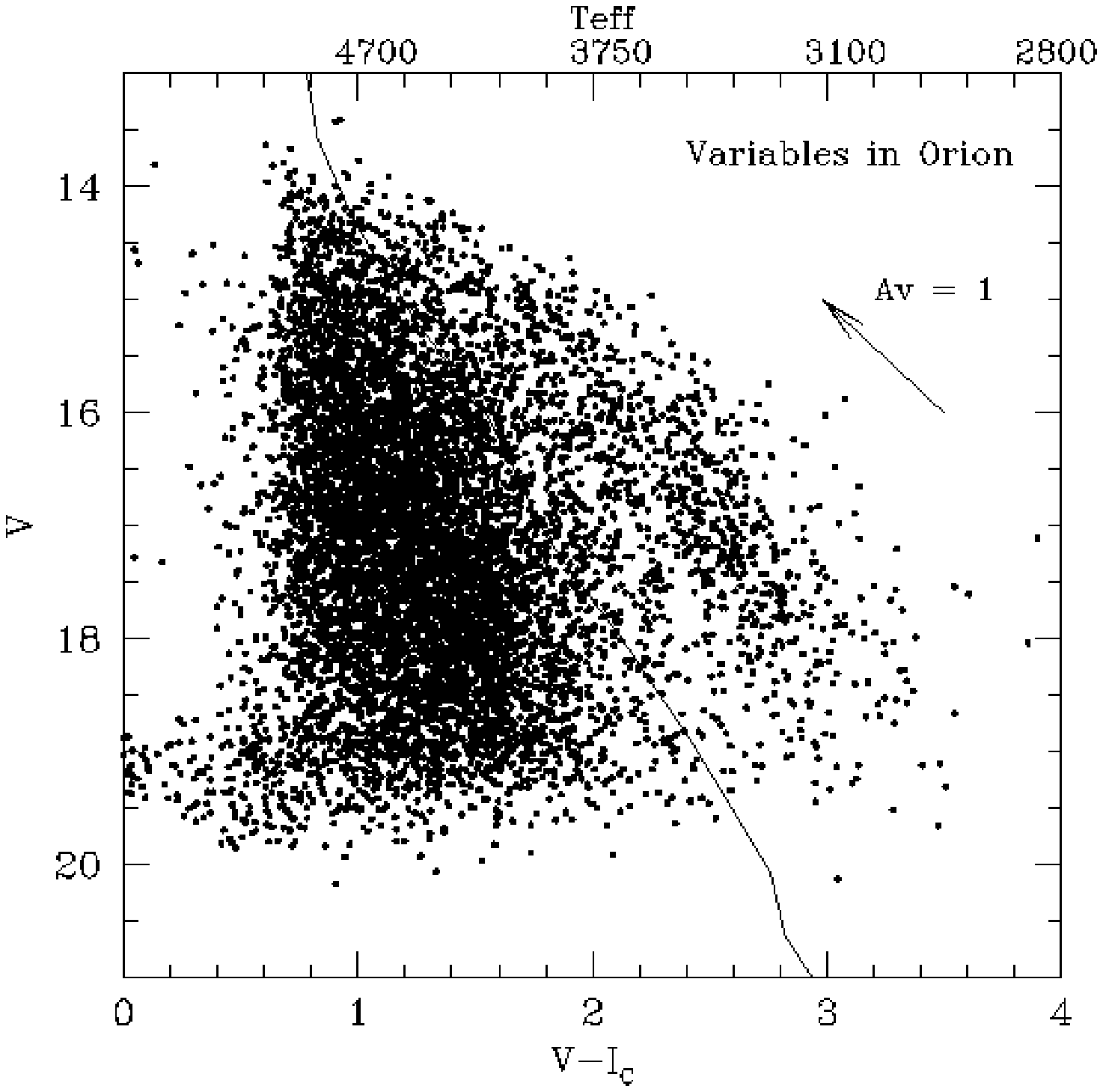}{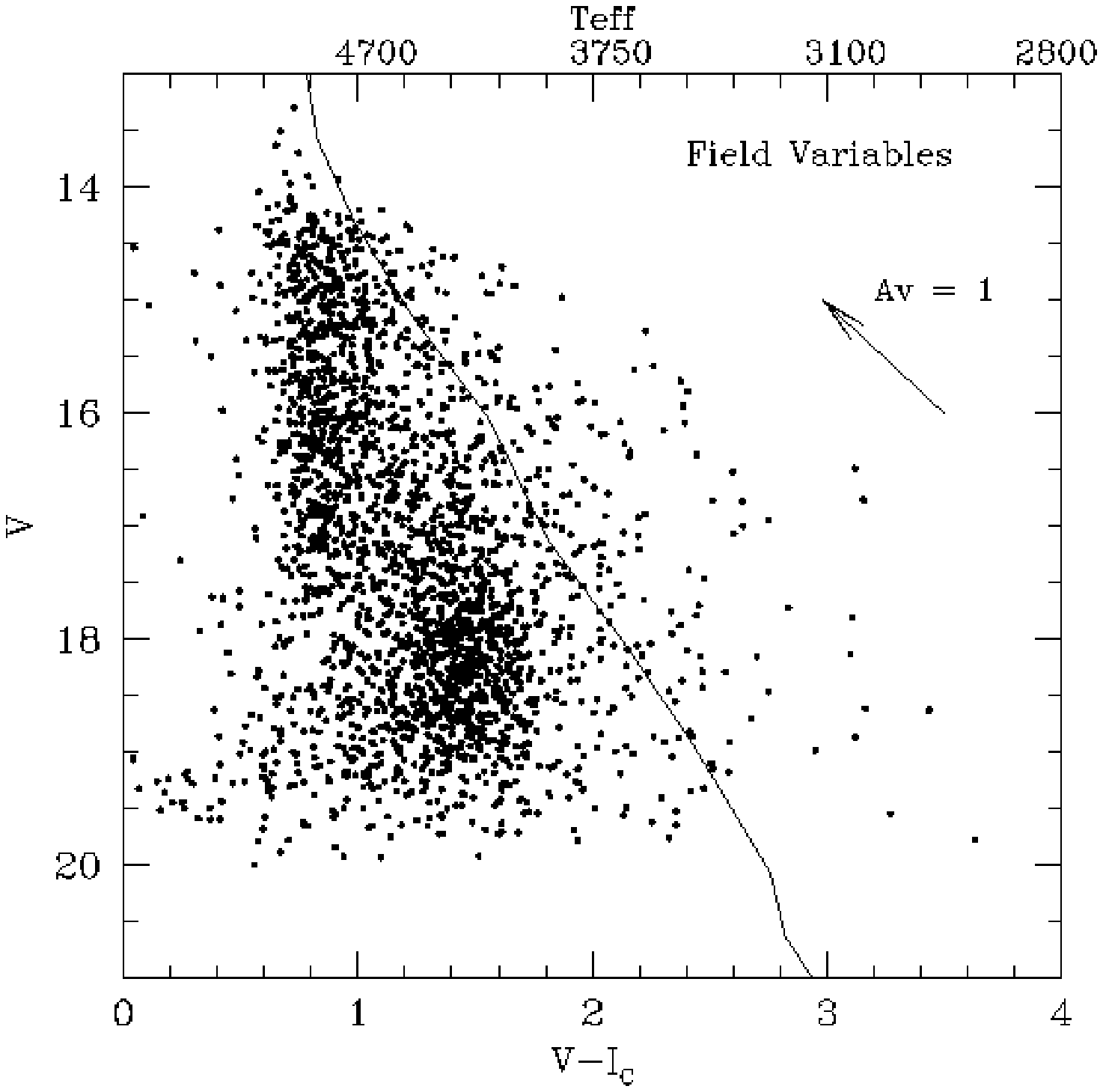}
\caption{Color-magnitude diagrams for stars selected as variable.
Left panel: all variables detected in the $\rm DEC=-1^\circ$ strip with $\rm RA > 5^h$.
Right panel: control field off Orion; stars in the same strip and with $\rm RA= 4^h48^m - 5^h00^m$.
}
\label{fig_varfield}
\end{figure}

In Figure 2 of \citet{bvc01} we showed that 
even in a modest region of $\rm \sim 10\> deg^2$ towards Ori OB1b there 
can be of the order of 4000 objects above the ZAMS down to $V\sim 19.5$;
after object lists are filtered by our variability
criterion only $\la 10$\% remain above the ZAMS.
Moreover, selection of variable stars above the ZAMS clearly
picks a significant fraction of the young stars in Orion, as
shown in Figure \ref{fig_varfield}, where we compare color-magnitude diagrams
for a region in OB1b ($l\sim 203^o$, $b\sim -18^o$) and a control field 
($l\sim 195^o$, $b\sim -31^o$), located well off
Orion; the Orion field exhibits an excess of objects above the
ZAMS not present in the control field, even after taking into account
that the Orion region has a higher density of stars
because it is closer to the galactic plane.
In the Orion field the fraction of variables above the ZAMS is
19\% compared to 10\% in the control field, a factor of $\sim 2$
over density. However, the densest
concentration of points in the color-magnitude diagram for the Orion field 
occurs at roughly 1 magnitude above the ZAMS. If we compare this
locus in both panels of Figure 2, the fraction of variables above the ZAMS
in the Orion field is a factor of 3 larger than in the control field.

We tested the reliability of the technique by recovering
as variables the 15 out of 19 previously known pre-main sequence
objects \citep{hbc88} among the stars selected as variable above
the ZAMS (see \S \ref{newmembers}). 
We also used our results to estimate the minimum
number of observations required to detect a star as variable.
We took the data for five of the \citet{hbc88} stars in our sample and
at random took out data points
from each star's light curve, running the $\chi^2$ test each time.
Accounting for points rejected due to objects falling on bad
columns, we find that a minimum of 5 observations are needed to
obtain a reliable detection.

The number of variables in the left panel of Figure \ref{fig_varfield}
decreases with increasing V and V-I.  This is a result both of the
decreased sensitivity in the V band for redder objects, and of
our selection criteria for variables. 
We selected a total of 3744 objects flagged as variable down to
our limit of V$\sim 19.7$,
and located above the ZAMS \citep{bca98} in the $V$ vs. $V-I_C$ diagram.
Within this sample we selected objects located above the 10 Myr isochrone
as our highest priority candidates 
(we excluded from this list the previously known T Tauri stars in the region).
We established a magnitude cutoff at $V=16$.
Brighter than this magnitude we included all variables above the ZAMS;
here we discuss follow up spectroscopy for 1083 of them.
 In the magnitude range $\rm V= 16 - 18.5$ we assigned the highest priority
to all variables above the ZAMS, we then added a number of non-variable
stars in the same region of the color-magnitude diagram. We adopted this
strategy in order to
circumvent the problem of missing a significant fraction of low-mass,
young stars at fainter magnitudes.
A total of 320 of the $\rm V= 16 - 18.5$ objects are discussed in this work.

Summarizing, our selection scheme of candidate Orion members started with 
20729 objects flagged as variables among 340532 detections. 
We then selected 3744 of the variables that were located above
the ZAMS in the color-magnitude diagram. 
Finally, we obtained spectra for a subset of those.
Here we discuss follow up spectroscopy for 1403 objects.

\subsection{Spectroscopy} 
\label{spectroscopy}

Follow up spectra are necessary to provide confirmation of pre-main sequence
low-mass stars. Optical spectra of TTS are characterized by emission in the
Balmer lines, particularly H$\alpha$, other lines like Ca I H \& K,
as well as the presence of the Li I$\lambda 6707${\AA } line in absorption.
\footnote{Because Li I is burned efficiently in the convective interiors of low mass
stars, the Li I line is a useful indicator of youth in mid-K to M type stars
with ages $\la 10$~Myr \citet{bhs97}.}  
Weak H$\alpha$ emission lines and Li I can be reliably measured in low
resolution spectra, which are easiest to obtain at many moderate
aperture telescopes \citep{bhs98,bvc01}.
Also, low resolution spectra usually span a large wavelength range, which
provides a reasonable number of features (like TiO absorption bands in
K through M type stars) for reliable spectral typing, from which 
$\rm T_{eff}$ can be derived.

\begin{figure}
\plotone{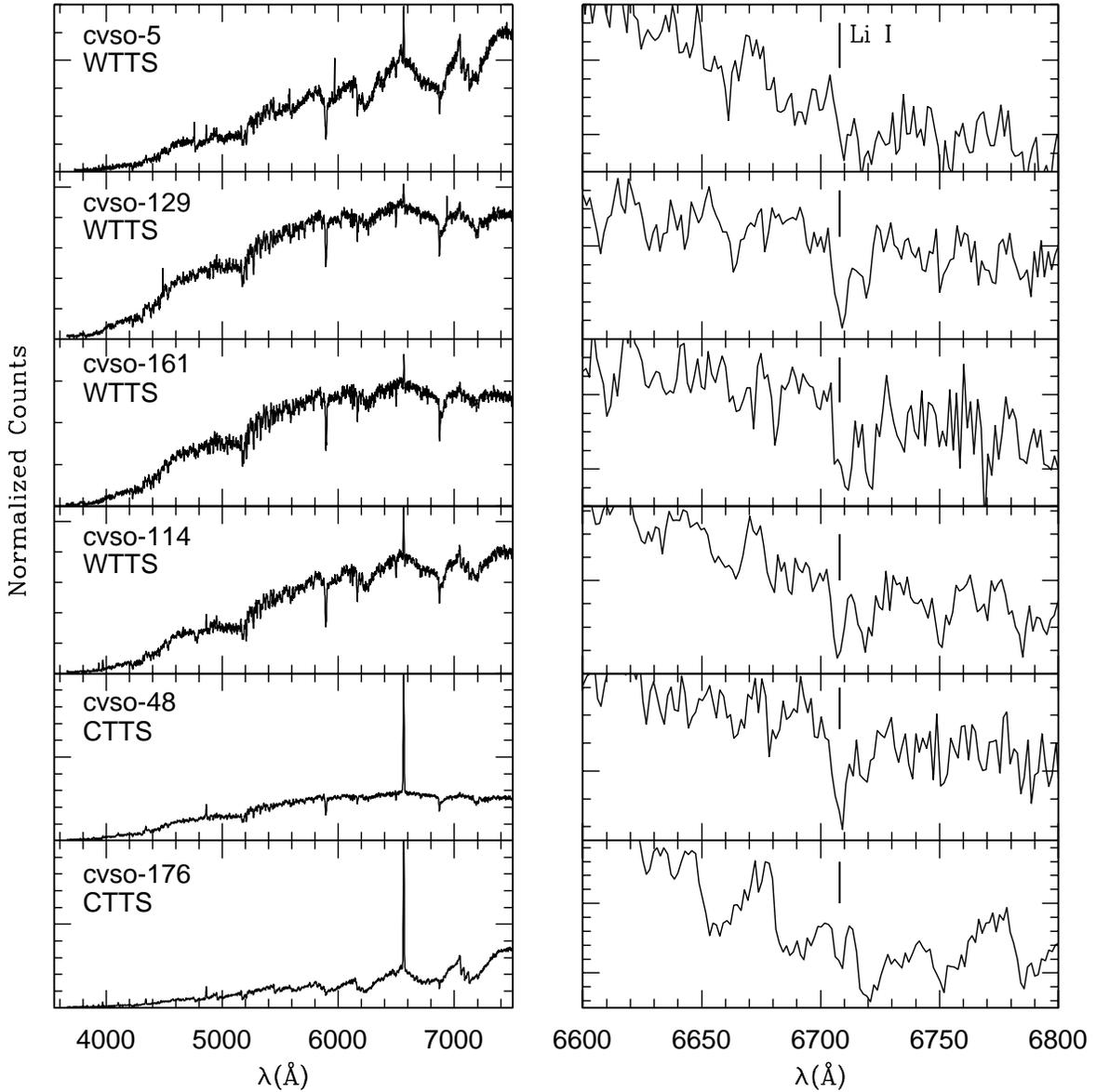}
\caption{Sample FAST spectra of newly identified low-mass young
stars in Orion. The left panels show the entire spectra from
$\sim 3800 - 7300$\AA, from weak H$\alpha$ emitting T Tauri stars
at top to Classical T Tauri stars, with their characteristic strong
emission in H$\alpha$, H$\beta$ and other lines, in the two lower left
panels. In the right panels we show an expanded view
of the wavelength range around the Li I line at 6707\AA; the
Ca I line next to it (6718\AA) is also clearly visible.
}
\label{fig_spectra}
\end{figure}

Optical spectra were obtained in queue mode 
for 1083 of the brighter candidates ($V < 16$)
during the period January 1999 through January 2002, 
using the 1.5 meter telescope of the
Whipple Observatory with the FAST Spectrograph \citep{fhs98}, 
equipped with the Loral $512 \times 2688$ CCD.
The spectrograph was set up in the standard configuration
used for ``FAST COMBO'' projects, a 300 groove $mm^{-1}$ grating and a 3''
wide slit. This combination offers 3400 {\AA } of spectral coverage centered
at 5500 {\AA}, with a resolution of 6 {\AA }.
The spectra were reduced at the CfA using software
developed specifically for FAST COMBO observations. All individual spectra
were wavelength calibrated using standard IRAF routines
The effective exposure times ranged from 60 s for the $V\sim 13$ stars
to $\sim 1500$ seconds for objects with $V\sim 16$. The
signal-to-noise ratio (SNR) of our spectra are typically $\ga 25$ at H$\alpha$,
sufficient for detecting equivalent widths down to a few tenths of \AA\ at our spectral
resolution.
In Figure \ref{fig_spectra} we show examples of typical FAST spectra.
 
We obtained optical spectra for a sample of faint targets ($V16 - 18.5$)
using the Hydra multi-object spectrograph installed on the KPNO 3.5m WIYN
telescope, during the nights of November 26 and 27, 2000 with clear weather.
We discuss here spectroscopy of 320 candidate objects distributed
in 5 fields spanning an area of $\rm \sim 4\Box^\circ$;
details are shown in Table \ref{hydralog}
(Note that Field 2b is the same region as Field 2a but with fibers located
on different sets of objects).
We used the Blue Channel fibers ($3\farcs 1$ diameter),
the Simmons Camera with the T2KC CCD, the 400@4.2 grating,
and G3\_GG-375 bench filter,
yielding a wavelength range $\sim 4000 - 7000${\AA } with a resolution
of 7.1\AA. 
All fields were observed with airmasses=1.0-1.5, and exposure
times for individual exposures were 1800 s. When weather allowed we obtained
two or three exposures per field.
Comparison CuAr lamps were obtained between each target field.

Following the strategy outlined in \S 2.4, 
in each Hydra field we first assigned fibers to all variables above the 
ZAMS and with V$ 16 - 18.5$, then filled the remaining fibers with non-variable objects
located in the pre-main sequence region of the $V$ vs. $V-I_C$ diagram
and with V$ 16 - 18.5$. 
When possible we also assigned fibers to bright $V < 16$
TTS confirmed from FAST spectra.
Typically we assigned 10-12 fibers to empty sky positions and
5-6 fibers to guide stars.
We used standard IRAF routines to remove the 
bias level from the two dimensional Hydra images.
Then the {\sl dohydra} package was used to extract
individual spectra, derive the wavelength calibration,
and do the sky background subtraction.
Since the majority of our fields are located in regions
with little nebulosity, background subtraction was in
general easily accomplished.

\section{Results}

\subsection{New members}
\label{newmembers}

Table \ref{tab_new} lists 197 new members of Ori OB1 found
in the survey so far.
(57 in Ori OB 1a, 138 in 1b, and 6 in 1c.) 
The first column of the
table gives the CIDA Variability Survey of Orion (CVSO) running number.
Column 2 gives the name of the counterpart
in the All Sky release of the 2MASS catalog
and 
columns 4 and 5 give the 2MASS positions.

The association between QuEST and 2MASS was made
as follows.
Positions obtained with the QuEST camera
are internally accurate to $0\farcs 2$. 
We used a 5\arcsec search radius to match our positions
for 165,530 objects in the $\rm DEC=-1^\circ$ strip
with 2MASS; we found a median offset of $0\farcs 67 \pm 0\farcs 22$.
\footnote{\citet{rma04} found a median offset of
$0\farcs 45 \pm 0\farcs 14$ when matching 198,213 objects
with the SDSS DR2.}
However, because the object surface density increases 
towards the galactic plane, confusion starts to be
an issue at $\rm RA \ga 5^h 30^m$; the automated pipeline
is not designed to deal with objects which are 
separated less than $\rm \la 2$ FWHM. The result is that
candidate lists have to be collated manually for blends.
Thus, we inspected visually the DSS red plate images for each
selected candidate. With a plate scale of $0\farcs 5$ the DSS can
resolve close pairs better than our QuEST camera images.
Still, some close pairs were observed spectroscopically (see below).
Once we determined the new members,
we searched in 
the 2MASS catalog for all sources within
10" from the QuEST position. 
We calculated the expected J magnitude
of the counterpart, from V and extinction
estimated from V-I using the interstellar
reddening law (\S \ref{properties}). For 191
stars, we found only one source
within this search radius with J within
0.5 magnitudes of the expected J. 
However, for 6 stars (2, 34, 59, 114, 170,
and 186 in Table \ref{tab_new}), we found two 2MASS sources with 
comparable J magnitudes within the search
radius. These stars were not resolved
in the QuEST photometry, shown in columns
5 and 6 of Table \ref{tab_new}, so we list
the table the two possible counterparts
and note that the optical photometry
corresponds to the blend of both sources.

In columns 10 and 11 of Table \ref{tab_new} we show
the equivalent width of
H$\alpha$ and Li 6707.
In column 12, we show the classification
of the star according to the equivalent
width of H$\alpha$. We follow \citet{wba03}
and use a limiting value between strong emission or Classical
T Tauri stars (CTTS, "C" in Table \ref{tab_new})
and weak T Tauri stars (WTTS, "W" in Table \ref{tab_new})
which depends on spectral type:
3 {\AA} if the spectral type is K5 or earlier,
10 {\AA} if the spectral type is between K5 and M2.5,
and 20 {\AA} if the spectral type is between M3 and M7.5.

\begin{figure}
\plotone{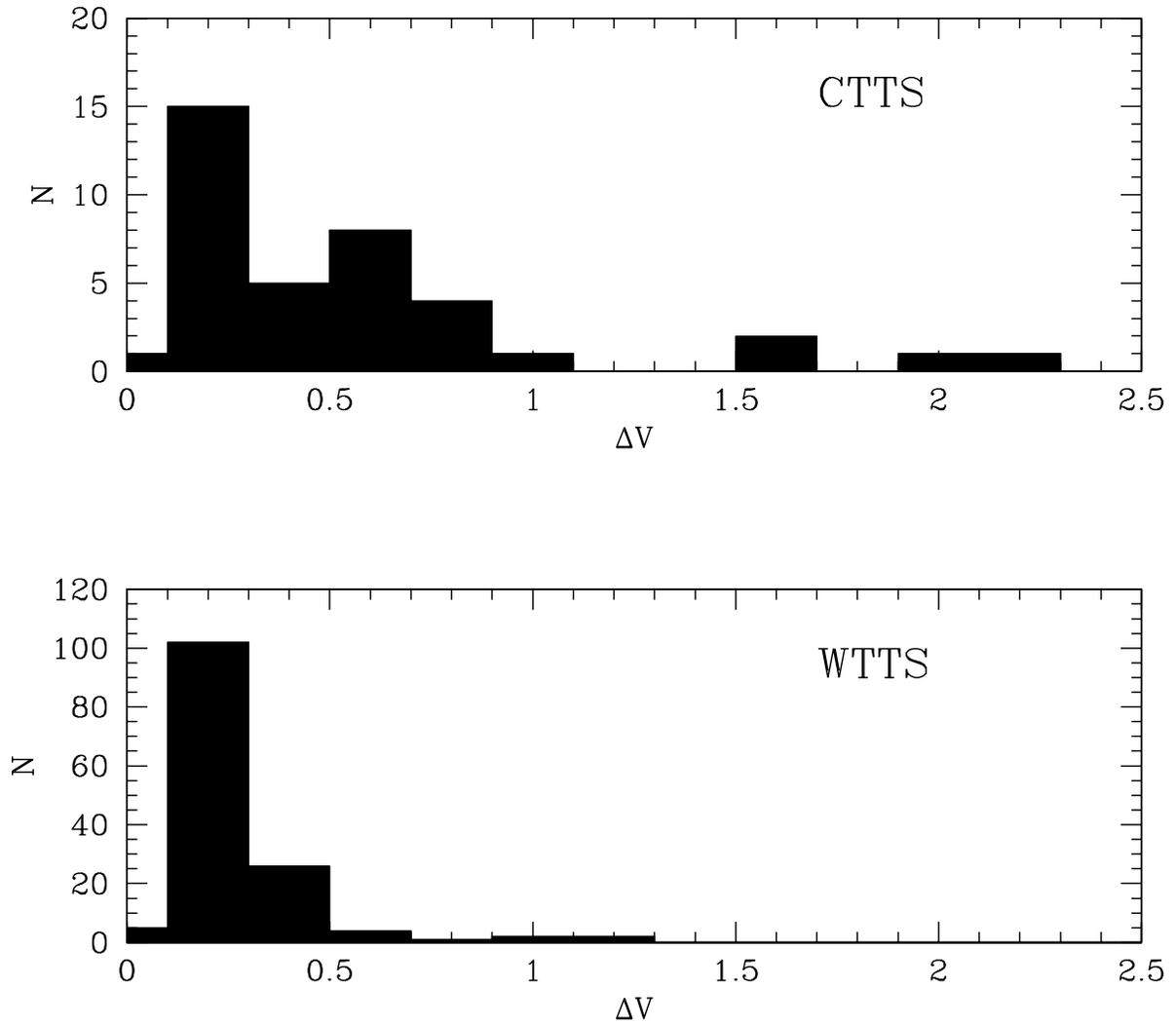}
\caption{Histograms showing the distribution of $\Delta {\rm mag}$ for
Classical T Tauri stars (upper panel) and Weak-lined T Tauri stars
(lower panel). 
}
\label{fig_amp}
\end{figure}

Figure \ref{fig_amp} shows the
distribution of ${\rm \Delta V}$ for
the stars in our sample. The average ${\rm <\Delta V>}$ for the
WTTS is 0.26 mag with a RMS=0.08 mag; for CTTS
we measure $<\Delta V> = 0.58$ mag with an average RMS=0.17 mag.
CTTS exhibit a wider range of $\Delta V$ with variations as 
large as $\sim 2$ magnitudes,
while the majority of WTTS vary less than 0.5 mag.
These values are consistent with those found
in the comprehensive study by \citet{hhg94}.
Columns 7 and 8 
list the standard deviation 
and the maximum ${\rm \Delta V}$, respectively.
Twenty-three stars do not have measurements for $\sigma (V)$
or $\Delta {V}$. These were objects located above the ZAMS
in the CMD but not detected as variable, to which we allocated
fibers in the WIYN-Hydra observations.

In columns 10 and 11 of Table \ref{tab_new} we show 
the equivalent width of
H$\alpha$ and Li 6707. 
In column 12, we show the classification
of the star according to the equivalent
width of H$\alpha$. We follow \citet{wba03}
and use a limiting value between Classical
T Tauri stars (CTTS, "C" in Table \ref{tab_new}) 
and Weak T Tauri stars (WTTS, "W" in Table \ref{tab_new})
which depends on spectral type. We use
the following limiting values for W(H$\alpha$):
3 {\AA} if the spectral type is K5 or earlier,
10 {\AA} if the spectral type is between K5 and M2.5,
and
20 {\AA} if the spectral type is between M3 and M7.5.
Unlike other regions of Orion, we are not
affected by nebular emission, so the
measurements of W(H$\alpha$) are fairly
reliable. Our main limitation is the low spectral
resolution, but even for the earlier type
stars in the survey the value of $\sim$ 3 {\AA}
is within our detection limit at our typical SNR.

Columns 14 to 16 list J and the colors
J-H and H-K for the 2MASS counterparts 
of each star. We give the near-IR magnitudes
of both possible counterparts for stars 
which were blended in the QuEST photometry.
The J magnitude of star 34
has a flag in the 2MASS catalog indicating
deblending problems; only the H-K color
is listed, for each of the counterparts,
and the actual magnitudes are given in
a note to the table. Similar deblending
problems are listed for J,H, and K of the
possible counterparts of star 186.

In Table \ref{tab_other} we indicate previously
existing identifications for each star (if any).
Many of these were identified in the Kiso survey for H$\alpha$ emitting stars.
This survey  contains 451 
objects in the region we have studied. Of these, 25\% are located
above the ZAMS and 50\% of this subset are flagged by our software as 
variable. Only 6\% of the Kiso objects below the ZAMS are variable.

There are 19 previously known T Tauri stars within our survey
area that have $V \ga 13.5$.
We detected 15 of them, all flagged as variable
(Table \ref{tab_known}); they are located
in the Lynds L1622, 1627 and 1630 clouds, associated with
to the embedded clusters NGC 2024, 2068 and 2071.
Stars LkH$\alpha$ 314, 315, 336/c and 337 were not detected
by the QuEST automated pipeline. In the case of 336/c, its
proximity ($5\arcsec$) to LkH$\alpha$ 336 results in the
images being too close to be singled as two distinct objects.
The remaining stars fell on bad columns in the reference
scan, so they were flagged as having problems by our software
and not included in the resulting object catalogs.

The depth ($\rm V\sim 19.7$) and wide spatial coverage of our
study offers several advantages when compared to
other large scale surveys aimed at finding pre-main sequence stars.
The ROSAT All-Sky Survey (RASS) was used to search for pre-main sequence
stars over $\rm 450\Box^o$ in Orion (Alcal\'a et al. [1996]).
However, in Brice\~no et al. (1997) we showed
that the shallow limit of the RASS results in significant
incompleteness for distances $d \ga 100$ pc, which
corresponds to $\rm V \sim 12$ for solar type stars with no extinction.
Deep X-ray observations can detect many young low-mass stars,
but they are limited to small selected fields
(e.g. the Chandra deep pointings in the ONC, Garmire et al. 2000).
The Kiso Schmidt H$\alpha$ survey covered a large area in Orion down
to $\rm V \sim 17$, but is biased toward the
CTTS and did not confirm membership of their sources.
Finally, while the 2MASS has being an important tool
to look for young stars embedded in their natal clouds, it is
not effective for finding the older, widespread populations we
are starting to unveil here.

\subsection{Spatial Distribution}
\label{espacial}

\begin{figure}
\plotone{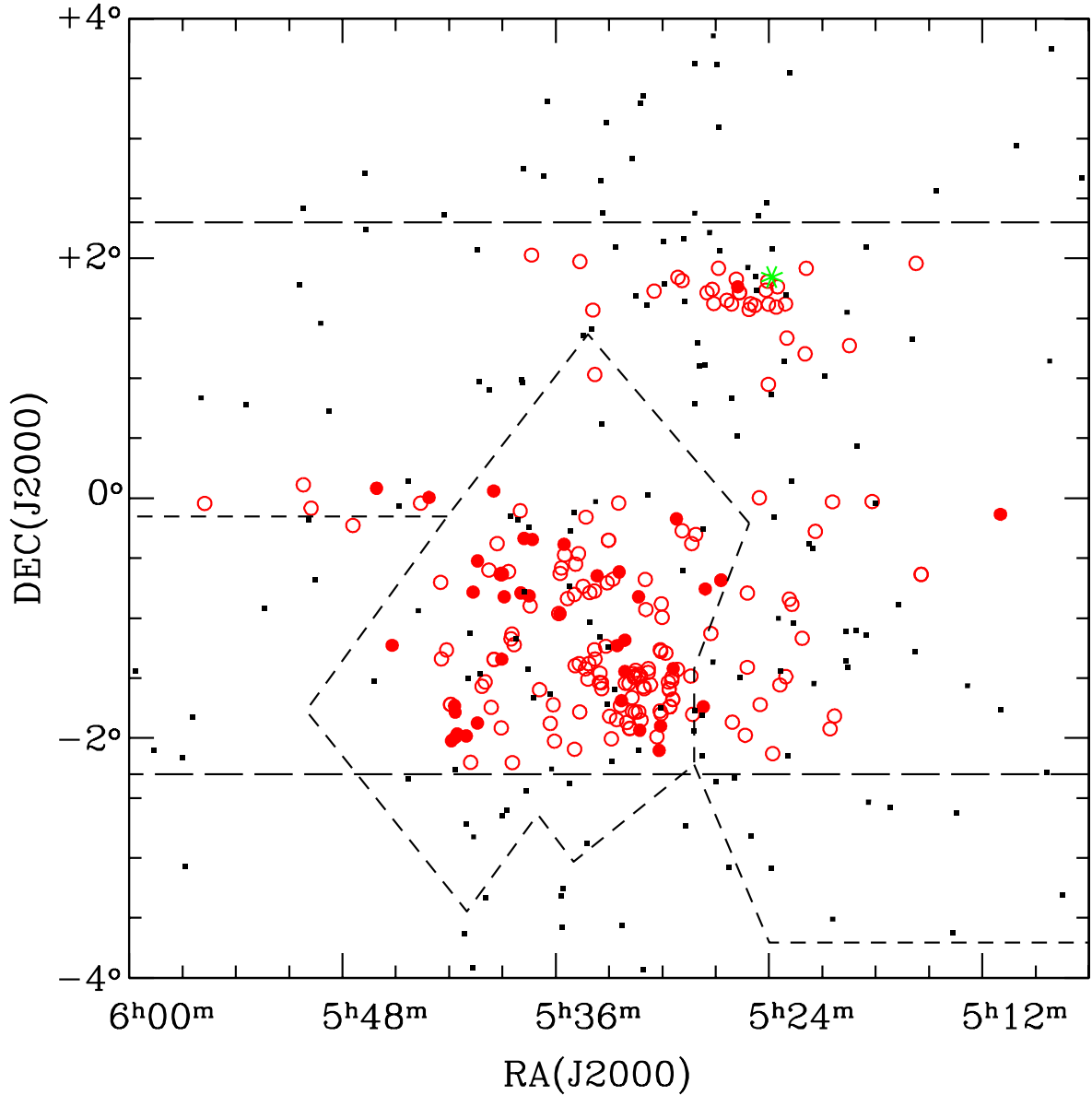}
\caption{Spatial distribution 
of the young stars found in the Orion survey:
CTTS (large solid circles), WTTS (large
open circles). The approximate limits of the scans are indicated
by long dashed. The small dashes line marks
the boundaries of Ori 1a, 1b, and 1c as
defined by \citet{wah77}. 
The small squares show the positions of the
OBA stars in the associations
(Hern\'andez et al. 2004),
selected from the Hipparcos catalog
with proper motions criteria from
from \citet{bgz94}. The asterisk marks
the Herbig Ae/Be star V346 Ori, located in the
clump of stars in Ori OB1a.
}
\label{spatialradec}
\end{figure}

\citet{bla64,bla91} was the first to
recognize the
the existence of different subassociations in
Orion, which he named 1a to 1d.
As inferred by the
hottest star in each subassociation, Blaauw assigned
an age of a few Myr to 1b and few $\times$ 10 Myr
to 1a,
and suggested that these groups
were one of the best examples of
sequential star formation. 
WH
carried out a detailed
photometric study of Blaauw's subassociations,
and were the first to formally assign
boundaries between them. 
Figure \ref{spatialradec} shows these
boundaries in the region of our survey.
We also show in this Figure
the distribution
of the new TTS stars, indicating
the CTTS and
WTTS with different symbols. It is apparent that 
the low mass stars are not distributed
uniformly throughout the survey area, rather they are
located in distinct regions; in particular,
the CTTS are concentrated
toward the 1b subassociation.

Figure \ref{spatialradec} also shows the distribution of the 
OB stars in the subassociations (Hern\'andez et al. 2004). 
These OB stars were selected from the 
Hipparcos catalog using proper motion criteria
given in Brown (1999, unpublished), in addition to photometric
and parallax limits (cf. Hern\'andez et al. 2004).
These stars
are indicated with small symbols in 
Figure \ref{spatialradec}. The low and high
mass stars appear to be similarly located 
in the sky.

One interesting structure is the grouping or ''clump''
of low-mass stars in 1a, located roughly at
$\rm RA=\> 5^h 23^m$
and $\rm DEC= \> 1^\circ 45'$. Its elongated form
in the E-W direction could be the result of this structure
lying at the northern limit of our $\rm DEC= +1^\circ$ strip.
Analysis of the $\rm DEC= +3^\circ$ strip (in a forthcoming
paper) will clarify whether it extends farther north.
This stellar group is clustered around the B1Vpe star 25 Ori,
a star with H$\alpha$ emission, but no
significant excess in the JHKs bands. However, the
A2 star V346 Ori (HIP 25299) is also in the cluster.
This star has a strong near-IR
excess and H$\alpha$ emission, for which it has been
classified  as a probable member of the
Herbig Ae/Be class (van den Ancker, de Winter, \& Tjin A
Djie 1998; Mora et al. 2001; Hern\'andez et al. 2004, in preparation).
The low mass members of the group are mostly
WTTS, with only one CTTS discovered so far, CVSO 35, which at a $\rm W(H\alpha)=-10.3\AA$
and a spectral type K7
is barely on the dividing line we have adopted for WTTS/CTTS.

WH and \citet{bgz94}
used main sequence fitting for the
OB stars in the subassociations
to estimate distances and ages.
They confirmed the age difference
indicated by the early studies of Blaauw.
Moreover, they found a difference
between 1a and 1b, with 1b having
a similar distance as that of the molecular cloud,
$\sim$ 440 pc \citep{grm81}, while
1a is closer, $\sim 330$ pc.
Stars in \citet{bgz94} were 
observed by Hipparcos; these results
were analyzed by \citet{bgz94},
who confirmed the distance difference
between 1a and 1b, 
although there were substantial observational errors
in the parallaxes of individual stars.

To reconsider the question of the
distance of 1a, we considered a subsample
of the Hipparcos stars close to the
clump of low mass stars in 1a
discussed above.
The average parallax of this subsample
is 3.1  with dispersion of 1.3 mas, 
compared to 
3.11 with dispersion of 1.9  mas for the much more spatially
spaced Brown's sample (Hern\'andez et al. 2004). The corresponding
mean distances are 323 +233/-96 pc and
322  +504/-122 pc so the distance of the
subsample is slightly better constrained.
We adopt this estimate
for the mean distance of 1a,
noting that the large dispersion
may suggest that the subassociation
have a significant depth, comparable to
the large spread in the sky.
This will result in a large
spread in the positions of the
stars in the color-magnitude (CMD) or
the H-R
 diagrams when using mean distances.

To further investigate the nature
of the subassociations, we 
compare the location of our new members
with the distribution of gas and dust
in the area. Figure \ref{galactic}
shows the stellar positions, together with
maps of $^{13}$CO from \citet{bally87},
which have enough resolution and probe
the densest gas regions. This
map is shown in color halftone scale.
We also
show the isocountour of color excess E($B-V)$ = 0.3
from \citet{sfdd98},
as indicator of the dust distribution
in the area. The gas and dust distributions
follow each other closely.

The boundaries of the survey observations
presented in this work are indicated, as
well as the WH boundaries. It can be
seen that Ori 1a corresponds generally to regions
devoid of gas and dust, outside the
boundaries of the Orion molecular clouds.
In contrast, Ori 1b is located at the
boundary of the clouds, in regions
of intermediate concentration
of gas and dust between 1a and the clouds.
In particular, a ring-like structure
with a radius of $\rm \sim 2^\circ$, which
correspond to $\sim$ 15 pc at the 1b
distance can be seen in low $^{13}$CO
emission and especially in the color
excess isocontour. 
The center of the ring is located
near the BOIae star $\epsilon$ Ori,
in a region of low $A_V$.

The part of the ring at $l \sim 203^\circ - 204^\circ$ falls on the
region of the $\rm DEC = +1^\circ$ 
observations, where
the sampling is poor because
spectroscopic follow-up 
was delayed by
the damaging of the some of the CCDs
in the QuEST camera (see \S \ref{sec_quest}).
The part of the ring at $l\ga 206^\circ$ falls outside
the survey limits presented in this paper.
However, the middle
part of the ring is very well sampled
by our observations at $\rm DEC = -1^\circ$.
A clear pattern can be seen in the
distribution of CTTS and WTTS. While
WTTS can be found in the entire
region, the CTTS concentrate towards
the regions of higher extinction and
gas density delineating the ring.
This distribution is similar to
that found in the molecular ring around
$\lambda$ Ori, which has comparable dimensions
to the Ori 1b ring ($\sim 16 - 20 $ pc, 
\citet[]{dom01,dom02}.
\citet{dom99}
found that only 5\% of the stars near the
center of the $\lambda$ Ori
ring are CTTS, while most CTTS concentrate near the
molecular ring.
These authors suggested that the
lack of accreting stars near the center
of the region
could be the result of photoevaporation
of disks by the nearby
OB stars. Dolan \& Mathieu propose
that the CTTS along the ring,
were formed by the effects of a
supernova at the center of the ring, 
which snow-plowed 
pre-existing molecular material into dense
enough concentrations to trigger star formation.
On a smaller scale, a similar situation is also
found by Sicilia-Aguilar et al. (2004)
in a study of the young cluster Trumpler 37,
where the CTTS are strongly concentrated
in a region away from the central O star.

We speculate that Ori OB 1b corresponds
to a population formed by a supernova event
which occurred near $\epsilon$ Ori $\sim$ 5 Myr
ago (\S \ref{sec_cmd}). The reality
of this scenario will be tested by forthcoming
radial velocity studies of the low mass 
population. Nonetheless, we have redefined the boundaries
of Ori 1b, so that it roughly follows the outer $A_V \sim$  0.5
contour at $l \sim 203 - 205^\circ$ and is limited
by the cloud at $l \ga 205^\circ$. This is the
definition used in Table \ref{tab_new}.
Obviously, whether a star belongs to 1a or 1b
near the boundary is very uncertain.

Finally, there are 6 stars in Table 3 which 
have $\rm RA\leq 5^h 44^m$ and $\rm DEC\sim 0^\circ$,
which places them around the boundary between the Ori OB 1a 
and OB 1c subassociations as defined by WH; they are
located on the molecular cloud. We have assigned them 
to Ori OB 1c.

\subsection{CMD diagrams}
\label {sec_cmd}

In Figures \ref{cmds} and \ref{cmdsbarafe}
we show the CMD diagram of Ori OB 1a and 1b,
together with theoretical isochrones from
\citet{sdf00} and \citet{bca98}.
Isochrones are represented assuming a 
distance of 330 pc for Ori OB 1a and 440 pc for Ori OB 1b.
Members are plotted with their observed magnitudes and
colors (not corrected for reddening). 

A large spread in ages is observed in
Figures \ref{cmds} and \ref{cmdsbarafe},
due in part to uncertainties in the distance
(\S \ref{espacial}). To restrict this somehow,
we plot in Figures \ref{cmds} and \ref{cmdsbarafe} (b)
only those stars located in the ''clump'' in 1a.
The spread is considerably reduced, showing
an age of 3 - 10 Myr for 1a. The spread
is much larger in 1b, where in addition
to the distance, we have uncertainties
due to lack of extinction corrections
and overlap with 1a.
In any event, it is
apparent that 1b has a much higher
fraction of 1-3 Myr stars than 1a.
We have included the Ori OB 1c stars together
with the Ori OB 1b members in the CMDs (labeled
as "1b+1c") because they are few objects and do not
impact our conclusions on the ages of each region.

\subsection{Stellar Properties}
\label{properties}

We have applied the spectral classification method
of \citet{hcb03}, extended to include
indices around the TiO bands for the late
spectral types of our stars. The resulting spectral
types are given in column 10 of Table \ref{tab_new}.
Errors in spectral
type are typically $\pm$ 1 subclass for
stars observed with the FAST spectrograph
and 2-3 subclasses
for stars observed with Hydra.
We have then obtained the effective temperature
T$_{eff}$ by using the calibration of
\citet[hereafter KH95]{kha95}.  
Using the V-I color, intrinsic colors
from KH95, and the \citet{ccm89}
extinction law with
R$_V$ = 3.1, we calculate the extinction $A_V$.
We dereddened the J magnitude and calculate
the stellar luminosity using the calibration
in KH95. These stellar properties are shown
in Tables \ref{tabpropob1a}, \ref{tabpropob1b}, and \ref{propob1c} for Ori 1a, 1b, and 1c,
respectively. 

The upper panels of Figures \ref{hrsiess} and \ref{hrbarafe}
show the location of stars in Ori 1a and 1b in the H-R diagram,
with evolutionary tracks and isochrones from 
\citet{sdf00}
\footnote{We calculate the effective temperature
from the luminosity and radius of each model,
to use the standard definition of this quantity.}. 
and
\citet{bca98}, respectively.
We calculate mass and ages using both
sets of tracks, and consider the
difference as the biggest uncertainty
in the determination of these quantities.
The derived quantities are given
in Tables \ref{tabpropob1a} to \ref{propob1c}.
As can be seen in Figure \ref{hrbarafe}, 
the Baraffe tracks cover a more restricted
range of parameter space than the Siess
tracks, so for a number of objects, only
the Siess values are given.

Although there is a large spread of the
stars in the H-R
 diagram, 
on average stars in 1a are older  
than in 1b. 
Table \ref{tab_ages} shows the average age and
standard distribution of the ages of 
1a and 1b.
An uncertainty of 100 pc
in the distance, results in an uncertainty
of $\sim 0.5$ in log L, which
in turn corresponds to a log age uncertainty of
$\sim 0.75 $, since $L \propto$ age$^-{2/3}$
\citep{har98}. The age difference 
between both associations is larger than
this uncertainty, in both determinations.

Stars in 1b show a larger spread
in the H-R
 diagram than those in 1a.
Since most of the stars
in Ori 1a are WTTS, any contamination
of stars of 1a into 1b would appear 
mostly in the WTTS. 
In Figures \ref{hrsiess} and \ref{hrbarafe}
we present H-R diagrams
for only the CTTS of 1b,
and in Table \ref{tab_ages} we show the average
age of 1b derived using only the CTTS.
Eliminating
the possible contamination of
1a stars by selecting only the
CTTS makes the age of 1b 2-3 Myr,
significantly younger than 1a.

In addition, we expect that stars with high values of $A_V$
are more likely to belong to 1b than to 1b.
We
show in the lower right panels of 
Figures \ref{hrbarafe}
and \ref{hrsiess} H-R diagrams
for all the CTTS and 
the WTTS which have $A_V> 0.4$.
The distribution of stars
in the H-R
 diagram is not significantly
different from that of the entire
sample, maybe reflecting a larger
range of ages in the WTTS than in the
CTTS possibly related to the formation
scenario discussed in 
\S \ref{espacial}.

\subsection{Near-IR excesses}

We have used the near-IR colors
in Table \ref{tab_new} to locate the stars 
in the JHK diagram, shown in Figure \ref{jhk}.
Also shown are the standard main sequence and giant
relations, 
the reddening vectors, and the CTTS locus,
the location of the stars surrounded
by accretion disks \citep{mch97}.
For Ori 1b, 
the WTTS concentrate around
the reddened standard sequence
up to early M, consistent
with the spectral type of the
sample and lack of reddening
toward 1a. Three out of the four
CTTS in 1a show infrared excesses.
We note that two of these stars,
40 and 41, are close to the
boundary 1a/1b and could very well
belong to 1b. However, star 35 with
H-K = 0.41, is a bona fide CTTS in 1a 
(see Table \ref{tab_new}). 

Much higher near-IR excesses are
found in 1b, as one would expect from the
the larger number of CTTS.
Figure \ref{jhk} shows
that stars with excesses 
generally are CTTS, consistent
with locations along the CTTS locus.

\section{Discussion and Conclusions}
\label{discussion}

A large fraction of the 
stars in our galaxy were formed in OB associations
\citep{wan00}. \citet{adm01} suggest
that many, if not most stars form in stellar groups 
with $10 < N < 100$ stars. 
These small groups are not strongly bound like
more massive entities destined to become open clusters,
and they evolve and disperse much more rapidly than clusters do.
Sparse, low density regions like Taurus and Chamaeleon are
more representative of what \citep{adm01} call isolated
star formation, while the $N\sim 100$ groups are expected
to form in OB associations. These groups are also expected
to consist mostly of solar type and less massive stars;
thus, they are not necessarily picked up by the existing
studies of massive stars in these regions.
Few studies exist that have been capable of mapping
the low mass, young population over 
wide enough areas in nearby OB associations (e.g. 
\citet{mml02}; \citet{pzg97}; 
\citet{dom02}).

We show here that multi-epoch surveys over very large areas
of the sky are powerful tools for identifying young, low-mass
stars in the older, extended regions of nearby OB associations 
were the molecular gas has dissipated and 
no longer serves as a marker of recent star formation.
These new type of surveys build on the availability of CCD mosaic
cameras on wide field telescopes of small to moderate size.

We have identified about 200 low mass ($\rm M \sim 0.12 - 0.9 \msun$)
pre-main sequence stars
widely spread over nearly $\rm 70\> deg^2$ in the Orion
OB1 association.
The spatial distribution of the newly identified young stars
indicate two large star forming events within this large area,
supporting previous studies that recognized two distinct
regions, the younger Ori OB1b and the older Ori OB1a. 
However, our sample is much more numerous, providing a better
estimate of the properties of each sub-association
 than studies of the more massive stars.
We find that 1b is almost completely confined to a large
bubble of gas of $\rm \sim 2^\circ$ radius roughly centered
on the BOIae star $\epsilon$ Ori.
We also note that there seems to be a rather well defined
edge to the OB association in the E-W direction, with
essentially no T Tauri stars found at $\rm RA \la 5^h 10^m$,
though many of our scans started further west than this.
Another striking result is the discovery of a concentration of stars
in 1a, located around an early type star,
which also includes a possible member of the
Herbig Ae/Be class.
This is similar, but
more numerous, than
other small clusters like $\eta$ Chamaleontis in the Sco-Cen
association \citep{lcm01}, possibly an example of the
grouped type of star formation proposed by \citep{adm01}.

Within present uncertainties (distances and depth of each
association, differences between theoretical evolutionary models, etc.) 
we find an age of $\sim 4-6$ Myr for 1b and 
$\sim 7-10$ Myr for 1a. The fact that almost all
accreting T Tauri stars (those with strong H$\alpha$ emission),
which are also the ones exhibiting near-IR excesses,
are concentrated in 1b, together with the environmental
characteristics of each region,
supports the idea of 1b 
being much younger than 1a, consistent with previous estimates
from the massive stars. 

In Figure \ref{diskfr} we plot
 the inner
 disk frequency as a function of age for
 a number of clusters and association, including
 our new data. The disk frequency has been determined
 by different methods. Haish, Lada, \& Lada (2001)
 measured the excess in the near-IR bands JHKL in a number
 of young clusters, spanning the age range 0.3 - 30 Myr.
 The disk frequency in Ori OB1a and 1b
 has been determined by the fraction of CTTS;
 these fractions are  $\sim$ 11\%
 and 23\%, respectively.
We estimate the fraction of
 accreting stars in the TW Hydra association,
using the breadth of the H$\alpha$ profile
from Muzerolle et al. (2000, 2001) as an indicator of accretion,
and adopting 24 as the total number of members
 from Song et al. (2003).
In general, there is a close correlation between the presence
of inner disks, as indicated by near-infrared excesses,
and the existence of accretion onto the central
star (Hartmann 1998).  While not all T Tauri inner disks exhibit
K-band excesses, depending upon inclination and magnetospheric
hole size (Meyer et al. 1997; Muzerolle et al. 2003), the
use of L-band data by Haisch \etal (2001)
should yield much more complete detections of disk excesses
(Hartigan \etal 1990).

The new data of Ori OB1a and 1b are
consistent with the
 previously determined decrease of the disk fraction
 with age (Hillenbrand \& Meyer 1999;
 Haisch \etal 2001) and with the
much more elaborate determination by Hillenbrand, Meyer, \& Carpenter (2004, in preparation),
supporting the idea of
 a timescale of $\sim$ 10 Myr for the dissipation
 of inner disks around low mass, young stars.

Finally, the discovery of a significant population of
low mass, pre-main sequence stars in a region devoid of gas
like 1a favors the growing general idea that molecular
clouds disperse over times of less than 10 Myr (Hartmann et al. 2001).

Acknowledgments. 
This work has been supported by
NSF grant AST-9987367 and NASA grant NAG5-10545.
C. Brice\~no acknowledges support from grant S1-2001001144 of
FONACIT, Venezuela.
This publication makes use of data products 
from the Two Micron All Sky Survey, which is a joint project 
of the University of Massachusetts and the Infrared 
Processing and Analysis Center/California Institute of 
Technology, funded by the National Aeronautics and Space 
Administration and the National Science Foundation.
This research has made use of the NASA/ IPAC Infrared Science Archive, 
which is operated by the Jet Propulsion Laboratory, California 
Institute of Technology, under contract with the National 
Aeronautics and Space Administration.
Grant.
We are grateful to Susan Tokarz at CfA, who is in charge of
the reduction and processing of FAST spectra.
We also thank Dr. Robert W. Wilson of the CfA Radio \& GeoAstronomy
Division for making available to us the $\rm ^{13}CO$ map shown
in Figure 6.
We thank the invaluable assistance of the observers and night
assistants at the Venezuela Schmidt telescope that made 
possible obtaining the data over these past years: 
A. Bongiovanni, C. Castillo, O. Contreras, D. Herrera,
M. Mart{\'\i}nez, F. Molina, F. Moreno, L. Nieves, G. Rojas, 
R. Rojas, L. Romero, U. S\'anchez, L. Torres
and K. Vieira. We also acknowledge the support from the
CIDA technical staff, and in particular of Gerardo S\'anchez.
Finally, we acknowledge the comments from an anonymous
referee, which helped improved this article.

\clearpage

\begin {deluxetable}{lcccccccc}
\tabletypesize{\scriptsize}
\tablewidth{0pt}
\tablecaption{CIDA 1m Schmidt Observing Log \label{tabcida}}
\tablehead{
\colhead{UT Date} & \multicolumn{4}{c}{Filters}& \colhead{Scan No.} & 
\colhead{RA(i)(J2000)} & \colhead{RA(f)(J2000)} & \colhead{DEC(J2000)} \\
\colhead{yyyy-mm-dd} & \multicolumn{4}{c}{} & \colhead{} & \colhead{(hh:mm)} & 
\colhead{(hh:mm)} & \colhead{(deg)} 
}
\startdata
1998-12-10 & R & B &  I & V &   501 & 04:10 & 05:51 & -1 \\
1998-12-11 & R & B &  I & V &   502 & 04:10 & 05:51 & -1 \\
1998-12-13 & R & B &  I & V &   507 & 04:10 & 05:51 & -1 \\
1998-12-13 & R & B &  I & V &   508 & 04:10 & 05:51 & -1 \\
1999-01-06 & R & B &  I & V &   511 & 03:55 & 05:53 & -1 \\
1999-01-06 & R & B &  I & V &   512 & 03:55 & 05:53 & -1 \\
1999-01-07 & R & B &  I & V &   513 & 03:55 & 05:53 & -1 \\
1999-01-08 & R & B &  I & V &   527 & 03:55 & 05:53 & -1 \\
1999-01-08 & R & B &  I & V &   528 & 03:55 & 05:53 & -1 \\
1999-01-09 & R & B &  I & V &   529 & 03:55 & 05:53 & -1 \\
1999-01-09 & R & B &  I & V &   530 & 03:55 & 05:53 & -1 \\
1999-01-21 & R & B &  I & V &   501 & 03:55 & 05:53 & -1 \\
1999-02-08 & R & B &  I & V &   501 & 03:55 & 05:53 & -1 \\
1999-03-27 & R & B &  R & V &   300 & 04:56 & 05:52 & -1 \\
1999-11-05 & R & Ha&  I & V &   550 & 03:55 & 05:51 & -1 \\
1999-11-05 & R & Ha&  I & V &   551 & 03:55 & 07:14 & -1 \\
1999-11-29 & R & Ha&  I & V &   506 & 04:15 & 07:59 & -1 \\
1999-11-30 & R & B &  I & V &   511 & 23:32 & 08:12 & -1 \\
1999-12-13 & U & B &  U & V &   510 & 03:05 & 09:10 & -1 \\
1999-12-14 & U & B &  U & V &   501 & 02:43 & 05:50 & -1 \\
1999-12-15 & U & B &  U & V &   501 & 23:53 & 06:13 & -1 \\
2000-01-06 & U & B &  U & V &   501 & 02:39 & 06:02 & -1 \\
2000-02-13 & R & B &  R & V &   601 & 03:23 & 05:21 & -1 \\
1999-12-16 & R & Ha&  I & V &   501 & 04:25 & 06:30 & +1 \\
1999-12-16 & R & Ha&  I & V &   502 & 04:26 & 06:30 & +1 \\
1999-12-17 & R & Ha&  I & V &   503 & 04:26 & 06:30 & +1 \\
1999-12-17 & R & Ha&  I & V &   504 & 04:25 & 06:30 & +1 \\
1999-12-26 & R & Ha&  I & V &   501 & 04:26 & 06:30 & +1 \\
1999-12-26 & R & Ha&  I & V &   502 & 04:26 & 06:31 & +1 \\
1999-12-27 & R & Ha&  I & V &   503 & 04:26 & 06:30 & +1 \\
1999-12-27 & R & Ha&  I & V &   504 & 04:26 & 06:30 & +1 \\
1999-12-27 & R & Ha&  I & V &   505 & 04:26 & 06:30 & +1 \\
1999-12-28 & R & Ha&  I & V &   506 & 04:26 & 06:30 & +1 \\
1999-12-29 & R & Ha&  I & V &   507 & 04:26 & 06:30 & +1 \\
1999-12-29 & R & Ha&  I & V &   508 & 04:26 & 06:30 & +1 \\
1999-12-29 & R & Ha&  I & V &   509 & 04:26 & 06:30 & +1 \\
2002-10-05 & V & R &  I & X &   510 & 04:49 & 06:12 & +1 \\
2002-10-13 & V & R &  I & X &   504 & 04:49 & 05:49 & +1 \\
2002-10-31 & V & I &  R & X &   503 & 04:49 & 06:12 & +1 \\
2002-10-31 & V & I &  R & X &   504 & 04:49 & 06:12 & +1 \\
2003-01-05 & V & I &  R & X &   501 & 04:49 & 06:10 & +1 \\
2003-01-05 & V & I &  R & X &   502 & 04:49 & 06:10 & +1 \\
2003-01-06 & V & I &  R & X &   501 & 04:49 & 06:12 & +1 \\
2003-01-06 & V & I &  R & X &   502 & 04:49 & 06:13 & +1 \\
2003-01-07 & V & I &  R & X &   500 & 04:49 & 06:11 & +1 \\
2003-01-07 & V & I &  R & X &   501 & 04:49 & 06:10 & +1 \\
2003-01-07 & V & I &  R & X &   502 & 04:49 & 06:12 & +1 \\
2003-01-08 & V & I &  R & X &   500 & 04:49 & 06:12 & +1 \\
2003-01-08 & V & I &  R & X &   501 & 04:49 & 05:58 & +1 \\
2003-01-08 & V & I &  R & X &   502 & 04:49 & 06:10 & +1 \\
2003-01-09 & V & I &  R & X &   500 & 04:49 & 06:10 & +1 \\
2003-01-28 & V & I &  R & X &   511 & 04:45 & 06:15 & +1 \\
2003-01-28 & V & I &  R & X &   512 & 04:45 & 06:15 & +1 \\
2003-02-01 & V & I &  R & X &   410 & 04:45 & 06:16 & +1 \\
2003-10-27 & V & I &  I & X &   501 & 05:10 & 05:56 & +1 \\
2003-10-27 & V & I &  I & X &   502 & 05:10 & 05:56 & +1 \\
\enddata
\tablenotetext{1}{Scan obtained for the Quest Equatorial Survey.}
\end{deluxetable}

\clearpage

\begin {deluxetable}{clcccc}
\tabletypesize{\scriptsize}
 \tablewidth{0pt}
 \tablecaption{WIYN Hydra Observing Log \label{hydralog}}
\tablehead{
\colhead{UT DATE} & \colhead{Field} & \colhead{RA(2000)} & \colhead{DEC(2000)} &
\colhead{Exptime} & \colhead{No. spectra} \\
\colhead{(yyyy mm dd)} & \colhead{} & \colhead{(hh:mm:ss)} & \colhead{($^\circ:\arcmin:\arcsec$)} &
\colhead{(sec)} & \colhead{}
}
\startdata
2000 11 27 & cbpms-04\_3   &  05:10:00.0  & -00:25:03 & 1800 &	56 \\
2000 11 26 & field1a	   &  05:28:33.1  & -00:39:00 & 1800 &	77 \\
2000 11 26 & field2a	   &  05:30:51.4  & -01:38:56 & 1800 &	70 \\
2000 11 27 & field2b	   &  05:30:51.4  & -01:38:56 & 1800 &	47 \\
2000 11 27 & field2b	   &  05:30:51.4  & -01:38:56 & 900  &  47 \\
2000 11 26 & cbpms-04\_10a &  05:38:00.0  & -00:25:03 & 1800 & 	70 \\
\enddata
\end{deluxetable}

\clearpage

\begin {deluxetable}{llllccrccrcccccc}
 \rotate
 \tabletypesize{\scriptsize}
 \tablewidth{0pt}
 \tablecaption{New Ori OB1 members\label{tab_new}}
 \tablehead{
\colhead{CVSO}& \colhead{2MASS} & \colhead{RA(2000)} & \colhead{DEC(2000)}& \colhead{V}& 
\colhead{V-I}& \colhead{$\rm \sigma(V)$} & \colhead{$\rm \Delta (V)$} & \colhead{SpT}& \colhead{$W$(H$\alpha$)}& \colhead{$W$(Li)}& 
\colhead{Cl}& \colhead{OB} & \colhead{J}&\colhead{J-H}&\colhead{H-K}\\
\colhead{}& \colhead{} & \colhead{hh:mm:ss} & \colhead{$\rm ^\circ \> \arcmin \> \arcsec$}& \colhead{}& 
\colhead{}& \colhead{} & \colhead{} & \colhead{}& \colhead{(\AA)}& \colhead{(\AA)}& 
\colhead{}& \colhead{} & \colhead{}&\colhead{}&\colhead{}
 }
\startdata
   1\tablenotemark{1}  &  05105877-0008030  &  05:10:58.78  &  -00:08:03.0  & 17.30 &  2.69 &  0.06 &  0.23 &  M3  & -37.80 &  0.40 &  C  &  1a  & 12.82 &  0.68 &  0.25 \\  
   2\tablenotemark{2}  &   05152615-0038067  &  05:15:26.15  &  -00:38:06.7  & 15.88 &  2.16 &  0.10 &  0.34 &  M1  &  -3.80 &  0.40 &  W  &  1a  & 12.55 &  0.67 &  0.23 \\  
      &   05152669-0038070  &  05:15:26.69  &  -00:38:07.0  & \nodata   &  \nodata  & \nodata   & \nodata   &  \nodata  &   &   &  \nodata  &  \nodata  & 12.81 &  0.63 &  0.25 \\  
   3  &    05154362+0157310  &  05:15:43.62  &  +01:57:31.1  & 16.25 &  2.25 &  0.03 &  0.10 &  M3  &  -5.90 &  0.30 &  W  &  1a  & 12.69 &  0.67 &  0.26 \\  
   4  &    05181171-0001356  &  05:18:11.71  &  -00:01:35.8  & 14.62 &  1.78 &  0.06 &  0.17 &  M0  &  -3.70 &  0.20 &  W  &  1a  & 11.72 &  0.67 &  0.20 \\  
   5  &    05192841+0116210  &  05:19:28.41  &  +01:16:21.2  & 15.45 &  2.10 &  0.04 &  0.16 &  M2  &  -5.10 &  0.70 &  W  &  1a  & 12.50 &  0.68 &  0.22 \\  
   6  &    05201885-0149010  &  05:20:18.85  &  -01:49:01.3  & 13.97 &  1.23 &  0.05 &  0.19 &  K5  &  -0.30 &  0.70 &
W  &  1a  & 11.66 &  0.60 &  0.20 \\
   7  &    05202587-0001490  &  05:20:25.87  &  -00:01:49.1  & 13.80 &  1.26 &  0.04 &  0.19 &  K6  &  -3.70 &  0.70 &
W  &  1a  & 11.14 &  0.64 &  0.16 \\
   8  &    05203382-0155237  &  05:20:33.82  &  -01:55:23.9  & 15.83 &  1.89 &  0.09 &  0.28 &  M0  &  -6.50 &  0.50 &
W  &  1a  & 12.35 &  0.71 &  0.22 \\
   9  &    05212330-0016361  &  05:21:23.30  &  -00:16:36.4  & 13.46 &  1.42 &  0.05 &  0.12 &  G5  &  -0.40 &  0.50 &
W  &  1a  & 11.28 &  0.68 &  0.16 \\
  10  &    05215424+0155020  &  05:21:54.26  &  +01:55:01.9  & 15.54 &  1.88 &  0.08 &  0.33 &  M0  &  -5.60 &  0.50 &
W  &  1a  & 12.31 &  0.68 &  0.24 \\
  11  &    05215832+0112140  &  05:21:58.32  &  +01:12:14.0  & 15.57 &  2.33 &  0.03 &  0.08 &  M2  &  -4.30 &  0.40 &
W  &  1a  & 12.10 &  0.66 &  0.23 \\
  12  &    05220790-0110059  &  05:22:07.90  &  -01:10:06.1  & 14.25 &  1.67 &  0.05 &  0.16 &  K7  &  -2.70 &  0.30 &
W  &  1a  & 11.55 &  0.68 &  0.23 \\
  13  &    05224282-0053065  &  05:22:42.82  &  -00:53:06.7  & 16.03 &  2.24 &  0.03 &  0.09 &  M2  &  -4.90 &  0.50 &
W  &  1a  & 12.63 &  0.65 &  0.24 \\
  14  &    05225160-0050322  &  05:22:51.61  &  -00:50:32.4  & 15.72 &  2.12 &  0.06 &  0.20 &  M2  &  -4.20 &  0.50 &
W  &  1a  & 12.48 &  0.60 &  0.25 \\
  15  &    05225965+0120089  &  05:22:59.66  &  +01:20:09.1  & 15.72 &  2.05 &  0.07 &  0.28 &  M1  &  -3.10 &  0.50 &
W  &  1a  & 12.57 &  0.74 &  0.19 \\
\enddata                                                
\tablenotetext{1}{Star observed with the Hydra spectrograph}
\tablenotetext{2}{Pair not resolved in Quest photometry.}
\tablenotetext{3}{H and K bands are inconsistently deblended}
\tablenotetext{4}{J band is inconsistently deblended. For 05415452-0143137, H=10.407 $\pm$  0.028
and K = 10.161  $\pm$ 0.032. For 05415479-0143131, H = 13.058 $\pm$ 0.057 and K = 12.328 $\pm$ 0.042} 
\tablecomments{Column (3) gives the 2MASS ID of the counterpart, located at a radius
given in column (4) in arcsec. In some cases (stars 34,114,186), more than one possible 
counterpart is shown (see text). Columns (5) and (6) give the 2MASS coordinates.
Spectral type errors are 1-2 subclasses for stars with FAST spectra 
and 2-3 subclasses for stars with Hydra spectra.
Equivalent widths $W$ in columns (10) and (11) are in {\AA}. 
The classification in column (12) (C=CTTS, W=WTTS) is based on spectral dependent
threshold $W$H($\alpha$) following White \& Basri (2003).
Assignation to a given subassociation in column (13) follows
discussion in \S \ref{espacial}.\\
}
\tablecomments{Table 3 is published in its entirety in the electronic version 
of the Astronomical Journal. A portion is shown here for guidance regarding 
its form and contents.}
\end{deluxetable}

\clearpage

\begin {deluxetable}{ll}
 \tabletypesize{\scriptsize}
 \tablewidth{0pt}
 \tablecaption{Other designation of new members\label{tab_other}}
 \tablehead{
\colhead{CVSO}& \colhead{Other Id}
}
\startdata
  40  &   Kiso A-0903 132                                \\  
  41  &   Kiso A-0903 133                                \\  
  47  &   Kiso A-0903 142                                \\  
  48  &   Kiso A-0903 143                                \\  
  58  &   Kiso A-0903 176                                \\  
  75  &   Kiso A-0903 191                                \\  
  82  &   IRAS 05284-0042                                \\  
  90  &   Kiso A-0904 1, Kiso A-0903 20                  \\  
 104  &   Haro 5-64, Kiso A-0903 220                     \\  
 107  &   Kiso A-0904 4                                  \\  
 109  &   V* V462 Ori, Haro 5-66, Kiso A-0904 5          \\  
 143  &   Kiso A-0904 36, SVS 1272, NSV 2328             \\  
 146  &   V* V499 Ori, Kiso A-0904 40, Haro 5-73, SVS 1238   \\  
 152  &   Haro 5-79                                      \\  
 155  &   Kiso A-0904 60                                 \\  
 157  &   Kiso A-0904 65                                 \\  
 164  &   Kiso A-0904 73                                 \\  
 165  &   Kiso A-0904 76                                 \\  
 171  &   Haro 5-87, Kiso A-0904 8                       \\  
 176  &   Haro 5-91, Kiso A-0904 104                     \\  
 177  &   Kiso A-0904 106                                \\  
 178  &   V* V513 Ori, Haro 5-92                         \\  
 180  &   Kiso A-0904 115                                \\  
 181  &   RX J054132.6-015658, [FS95] 14                 \\  
 182  &   Haro 5-59, NGC 2024 HLP 4                      \\  
 184  &   LkH$\alpha$ 288                                \\  
 187  &   Kiso A-0904 131                                \\  
 190  &   V* V518 Ori, Haro 5-93, Kiso A-0904 148        \\  
 192  &   LkH$\alpha$ 293, Haro 5-95, Kiso A-0904 165    \\  
 193  &   LkH$\alpha$ 298, GSC 00116-01169, CSI+00-05435 \\  
      &   Kiso A-0904 176, RJHA 11, SSV LDN 1630 51    \\  
\enddata
\end{deluxetable}

\begin {deluxetable}{clcccclrl}
 \tabletypesize{\scriptsize}
 \tablewidth{0pt}
 \tablecaption{Previously known T Tauri stars within the survey area \label{tab_known}.}
 \tablehead{
\colhead{HBC No.} & \colhead{Designation} & \colhead{RA(J2000)} & \colhead{DEC(J2000)} & \colhead{$V$} & 
\colhead{$V-I_C$} & \colhead{SpT} & \colhead{W(H$\alpha$)} & \colhead{Location} \\
\colhead{}& \colhead{} & \colhead{hh:mm:ss} & \colhead{$\rm ^\circ \> \arcmin \> \arcsec$}& \colhead{(mag)} & 
\colhead{(mag)} & \colhead{} & \colhead{(\AA)} & \colhead{} 
 }
\startdata
 161 & PU Ori               &  05:36:23.71 & -00:42:32.9 &  15.043  & 1.031  & K3:   &    42.0 &   \\
 501 & CoKu NGC 2068/1      &  05:44:33.86 & -01:21:36.5 &  17.412  & 3.131  & K7,M0 &     9.0 &   L1630 \\
 503 & LkH$\alpha$ 301      &  05:46:19.47 & -00:05:19.6 &  14.694  & 2.129  & K2    &    12.0 &   L1627 \\
 504 & LkH$\alpha$ 302      &  05:46:22.45 & -00:08:52.6 &  15.106  & 1.976  & K3    &    23.0 &   L1627 \\
 505 & LkH$\alpha$ 307      &  05:47:06.01 & +00:32:08.9 &  17.243  & 3.250  & cont  &   154.0 &   L1630 \\
 506 & LkH$\alpha$ 309      &  05:47:06.98 & +00:00:47.6 &  14.766  & 2.516  & K7    &    51.0 &   L1627 \\
 507 & LkH$\alpha$ 313      &  05:47:13.95 & +00:00:17.1 &  15.791  & 1.633  & M0.5  &    20.0 &   L1627 \\
 508 & LkH$\alpha$ 312      &  05:47:13.97 & +00:00:16.5 &  14.911  & 1.933  & M0    &     8.2 &   L1627 \\
 510 & LkH$\alpha$ 316      &  05:47:35.73 & +00:38:40.9 &  14.357  & 1.927  & cont  &    79.0 &   L1630 \\
 511 & LkH$\alpha$ 316/c    &  05:47:36.79 & +00:39:14.7 &  16.777  & 3.220  & M0.5  &     2.9 &   L1630 \\
 512 & LkH$\alpha$ 319      &  05:47:48.37 & +00:40:59.3 &  14.639  & 1.822  & M0    &    61.0 &   L1630 \\
 513 & LkH$\alpha$ 320      &  05:48:01.07 & +00:34:30.7 &  14.730  & 2.116  & M1    &    16.0 &   L1630 \\
 188 & LkH$\alpha$ 334      &  05:53:40.91 & +01:38:14.1 &  13.864  & 1.224  & K4    &    23.0 &   L1622 \\
 189 & LkH$\alpha$ 335      &  05:53:58.69 & +01:44:09.5 &  14.023  & 1.601  & K4    &    47.0 &   L1622 \\
 190 & LkH$\alpha$ 336      &  05:54:20.12 & +01:42:56.5 &  14.186  & 1.696  & K7,M0 &    24.0 &   L1622 \\
\enddata
\tablecomments{Positions and optical photometry from our data. 
Spectral types, W(H$\alpha$) and molecular cloud 
were the star is located are from the \citet{hbc88} catalog.}
 \end{deluxetable}

\clearpage

\begin {deluxetable}{lccccccccc}
 \tabletypesize{\scriptsize}
 \tablewidth{0pt}
 \tablecaption{Properties of Ori OB1a stars \label{tabpropob1a}}
 \tablehead{
\colhead{CVSO}&  \colhead{T$_{eff}$}&\colhead{$A_V$}& \colhead{L} & \colhead{R} &\colhead{M(B)} & \colhead{age(B)}& \colhead{M(S)}&
\colhead{age(S)} & \colhead{Cl}\\
\colhead{}& \colhead{K}&\colhead{mag}& \colhead{$\lsun$} & \colhead{$\rsun$} &\colhead{$\msun$} & \colhead{Myr}& \colhead{$\msun$}&
\colhead{Myr} & \colhead{}\\
 }
\startdata
   1&  3470.&   0.53&   0.16&   1.12&   0.42&   4.60&   0.33&   4.53&CTTS\\
   2\tablenotemark{1}&  3720.&   0.48&   0.22&   1.14&   0.65&   7.47&   0.47&   6.07&WTTS\\
   &  3720.&   0.48&   0.18&   1.01&   0.65&  10.81&   0.47&   7.97&WTTS\\
   3&  3470.&   0.00&   0.16&   1.11&   0.42&   4.73&   0.33&   4.68&WTTS\\
   4&  3850.&   0.00&   0.43&   1.48&   0.72&   3.25&   0.57&   3.06&WTTS\\
   5&  3580.&   0.00&   0.20&   1.17&   0.55&   5.96&   0.38&   4.54&WTTS\\
   6&  4350.&   0.00&   0.49&   1.23& \nodata& \nodata&   0.98&  13.10&WTTS\\
   7&  4205.&   0.00&   0.78&   1.66& \nodata& \nodata&   0.93&   4.25&WTTS\\
   8&  3850.&   0.22&   0.26&   1.14&   0.73&   7.72&   0.58&   7.64&WTTS\\
   9&  5770.&   1.62&   1.21&   1.10& \nodata& \nodata&   1.13&  94.56&WTTS\\
  10&  3850.&   0.19&   0.26&   1.15&   0.73&   7.36&   0.58&   7.37&WTTS\\
  11&  3580.&   0.46&   0.33&   1.49&   0.57&   3.12&   0.39&   2.51&WTTS\\
  12&  4060.&   0.17&   0.55&   1.50&   0.82&   3.31&   0.78&   4.99&WTTS\\
  13&  3580.&   0.24&   0.19&   1.14&   0.55&   6.33&   0.38&   4.94&WTTS\\
  14&  3580.&   0.00&   0.21&   1.18&   0.55&   5.83&   0.38&   4.41&WTTS\\
  15&  3720.&   0.22&   0.21&   1.09&   0.65&   8.54&   0.47&   6.78&WTTS\\
\enddata
\tablenotetext{1} {Properties have been determined
for each of the two 2MASS counterparts using
the corresponding J magnitude}.
\tablecomments{``B'' and ``S'' refer to quantities
determined using the \citet{bca98} and \citet{sdf00}
evolutionary tracks}
\tablecomments{Table 6 is published in its entirety in the electronic version 
of the Astronomical Journal. A portion is shown here for guidance regarding 
its form and contents.}
 \end{deluxetable}

\begin {deluxetable}{lccccccccc}
 \tabletypesize{\scriptsize}
 \tablewidth{0pt}
 \tablecaption{Properties of Ori OB1b stars \label{tabpropob1b}}
 \tablehead{
\colhead{CVSO}& \colhead{T$_{eff}$}&\colhead{$A_V$}& \colhead{L} & \colhead{R} &\colhead{M(B)} & \colhead{age(B)}& \colhead{M(S)}&
\colhead{age(S)} & \colhead{Cl}\\
\colhead{}& \colhead{K}&\colhead{mag}& \colhead{$\lsun$} & \colhead{$\rsun$} &\colhead{$\msun$} & \colhead{Myr}& \colhead{$\msun$}&
\colhead{Myr} & \colhead{}\\
 }
\startdata
  37&  3850.&   0.00&   0.41&   1.43&   0.72&   3.62&   0.58&   3.59&WTTS\\
  48&  4590.&   0.83&   1.23&   1.76& \nodata& \nodata&   1.28&   6.71&CTTS\\
  50&  4205.&   0.00&   1.51&   2.31& \nodata& \nodata&   0.91&   1.68&WTTS\\
  52&  4060.&   0.00&   0.40&   1.28&   0.86&   6.01&   0.80&   8.10&WTTS\\
  56&  4060.&   0.00&   0.47&   1.38&   0.84&   4.49&   0.80&   6.61&WTTS\\
  58&  4060.&   0.00&   0.57&   1.52&   0.82&   3.16&   0.78&   4.77&CTTS\\
  59\tablenotemark{1}&  3470.&   0.65&   0.14&   1.05&   0.42&   5.55&   0.32&   5.46&WTTS\\
  &  3470.&   0.65&   0.48&   1.92&   0.52&   1.27&   0.34&   1.66&WTTS\\
  60&  3470.&   0.36&   0.11&   0.90&   0.42&   8.61&   0.31&   7.53&WTTS\\
  61&  3720.&   1.22&   0.13&   0.85&   0.65&  18.41&   0.46&  12.20&WTTS\\
  62&  4590.&   1.07&   1.77&   2.10& \nodata& \nodata&   1.41&   3.82&WTTS\\
  63&  4590.&   0.32&   1.44&   1.90& \nodata& \nodata&   1.32&   5.46&WTTS\\
  64&  3470.&   0.36&   0.10&   0.88&   0.42&   9.08&   0.31&   7.77&WTTS\\
  65&  4205.&   0.00&   0.47&   1.29& \nodata& \nodata&   0.94&   9.14&WTTS\\
  66&  3720.&   0.55&   0.28&   1.26&   0.66&   5.50&   0.47&   4.41&WTTS\\
\enddata
\tablenotetext{1} {Properties have been determined
for each of the two 2MASS counterparts using
the corresponding J magnitude}.
\tablecomments{``B'' and ``S'' refer to quantities
determined using the \citet{bca98} and \citet{sdf00}
evolutionary tracks}
\tablecomments{Table 7 is published in its entirety in the electronic version 
of the Astronomical Journal. A portion is shown here for guidance regarding 
its form and contents.}
 \end{deluxetable}

\clearpage

\begin {deluxetable}{lccccccccc}
 \tablewidth{0pt}
 \tablecaption{Properties of Ori OB1c stars \label{propob1c}}
 \tablehead{
\colhead{CVSO}&  \colhead{T$_{eff}$}&\colhead{$A_V$}& \colhead{L} & \colhead{R} &\colhead{M(B)} & \colhead{age(B)}& \colhead{M(S)}&
\colhead{age(S)} & \colhead{Cl}\\
\colhead{}& \colhead{K}&\colhead{mag}& \colhead{$\lsun$} & \colhead{$\rsun$} &\colhead{$\msun$} & \colhead{Myr}& \colhead{$\msun$}&
\colhead{Myr} & \colhead{}\\
 }
\startdata
 192&  4205.&   1.95&   1.53&   2.33& \nodata& \nodata&   0.91&   1.65&CTTS\\
 193&  4205.&   1.21&   1.53&   2.33& \nodata& \nodata&   0.91&   1.64&CTTS\\
 194&  4730.&   0.00&   0.70&   1.25& \nodata& \nodata&   1.05&  18.70&WTTS\\
 195&  4730.&   0.42&   0.44&   0.99& \nodata& \nodata&   0.92&  29.70&WTTS\\
 196&  4205.&   0.00&   0.42&   1.21& \nodata& \nodata&   0.93&  10.83&WTTS\\
 197&  5800.&   2.32&   0.86&   0.92& \nodata& \nodata& \nodata& \nodata&WTTS\\
\enddata
\tablenotetext{1} {Properties have been determined
for each of the two 2MASS counterparts using
the corresponding J magnitude}.
\tablecomments{``B'' and ``S'' refer to quantities
determined using the \citet{bca98} and \citet{sdf00}
evolutionary tracks}
 \end{deluxetable}

\clearpage
\begin {deluxetable}{lcc}
 \tablewidth{0pt}
 \tablecaption{Ages of OB 1a and 1b \label{tab_ages}}
 \tablehead{
\colhead{OB} & \colhead{age(B)}& \colhead{age(S)} \\
\colhead{}&  \colhead{Myr}& \colhead{Myr} \\
 }
\startdata
1a  &   7.4  $\pm$   0.6 & 10.7$\pm$  2.0 \\
1b  &   6.0    $\pm$ 0.8 & 5.3  $\pm$0.4 \\
1b CTTS & 3.6    $\pm$  1.0& 3.6 $\pm$ 0.5 \\
\enddata
\tablecomments{``B'' and ``S'' refer to quantities
determined using the \citet{bca98} and \citet{sdf00}
evolutionary tracks}
 \end{deluxetable}

\clearpage

\clearpage

\clearpage

\clearpage

\clearpage

\clearpage
{\rotate
\begin{figure}
\hskip -1.8in
\includegraphics[scale=0.75,angle=270,keepaspectratio=true]{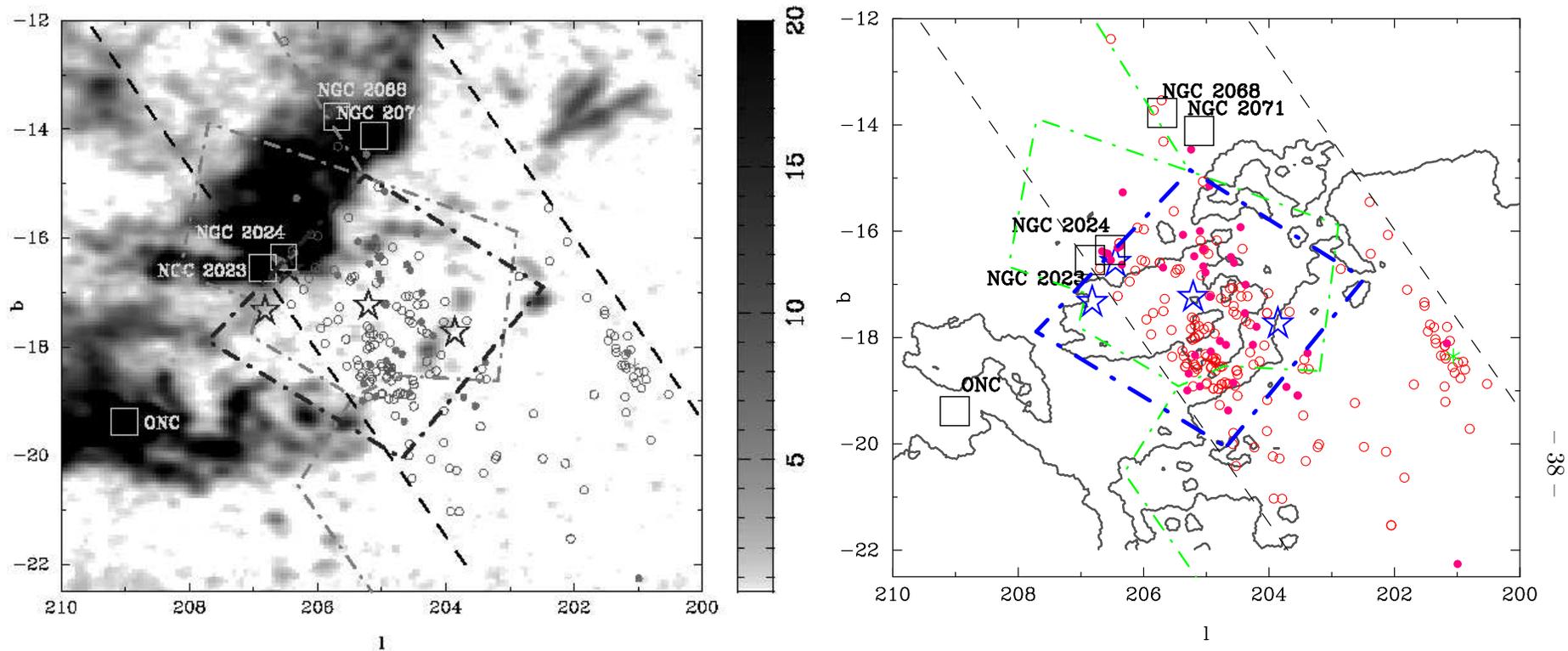}
\vskip -4.0in
\hskip 3.4in
\includegraphics[scale=0.62,angle=0,keepaspectratio=true]{Briceno.fig6b.ps}
\caption{Spatial distribution of the new
stars in galactic coordinates. 
Solid dots show the CTTS and open circles
show the WTTS. 
The location
of the ONC, NGC 2023, and NGC 2024 (open
rectangles), 
and the stars in the belt
(large blue stars) is indicated.
The halftone color map in the left panel shows the
integrated  $^{13}$CO emissivity
from \citet{bally87} covering the range from
from 0.5 to 20 ${\rm K \, km \, s^{-1}}$.
The isocontour 
for $A_V = $ 1 from Schlegel et al. (1998) is plotted
in the right panel. 
In dot-dash green lines we show
the \citet{wah77} boundaries, and in thick blue dot-dash
lines the boundaries of Ori OB1b adopted in
this work. The Herbig Ae/Be star V346 Ori is
shown as a green star.
The approximate limits of the scans are indicated
by long dashed.
}
\label{galactic}
\end{figure}
}

\clearpage
\begin{figure}
\plotone{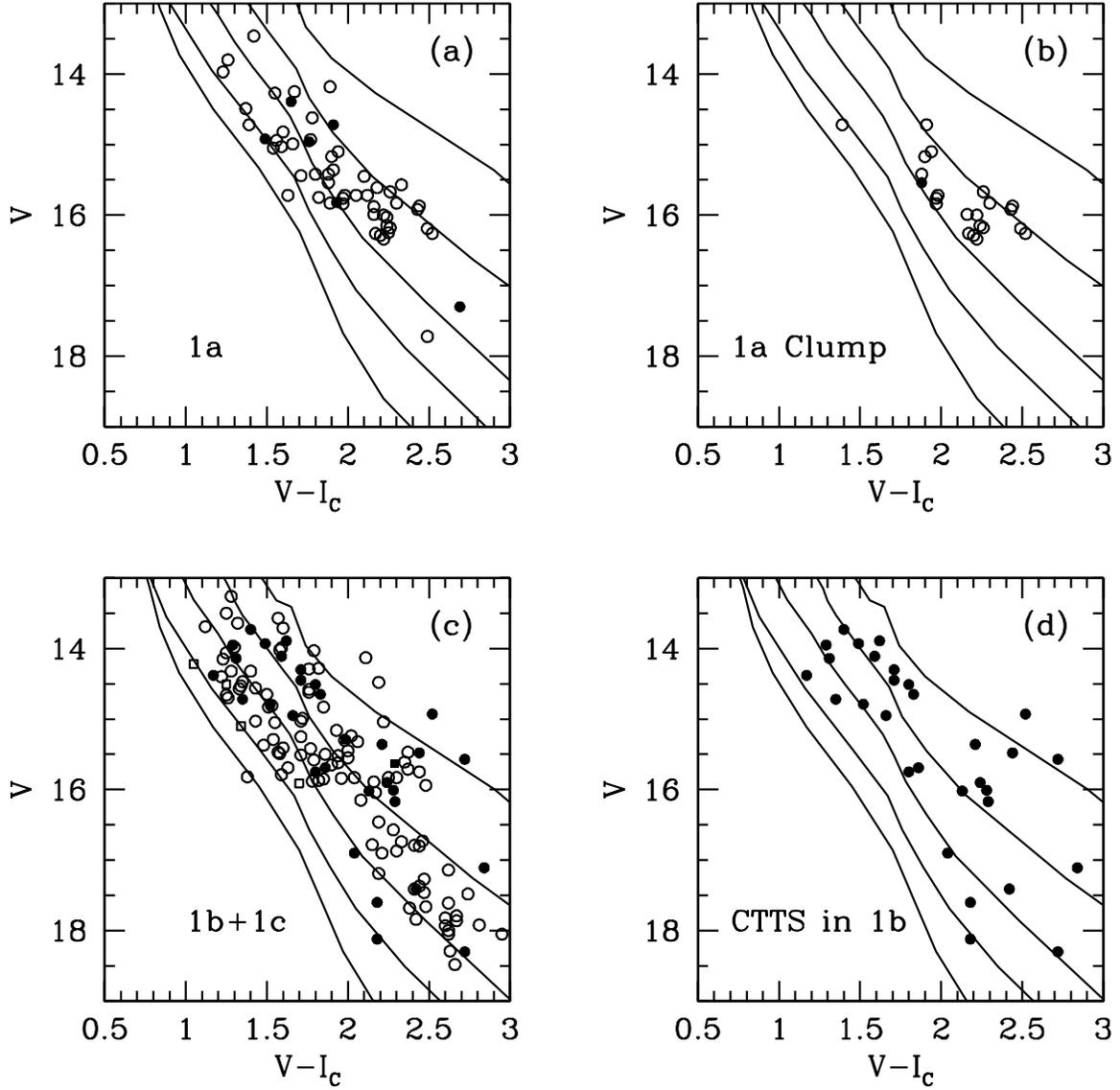}
\caption{Observed color-magnitude diagrams for the subassociations
superimposed on the \citet{sdf00} isochrones, ordered
from top to bottom as 1 Myr, 3 Myr, 10 Myr, and 30 Myr.
The lower line is the ZAMS.
CTTS (solid symbols) and WTTS (open symbols) are indicated. (a) Ori 1a,
(b) Clump in Ori 1a, (c) Ori 1b (circles) and 1c (squares), and (d) 
Only CTTS in Ori 1b. 
Distances of 330 pc and 440 pc have been used for
Ori 1a and 1b, respectively.
}
\label{cmds}
\end{figure}

\clearpage
\begin{figure}
\plotone{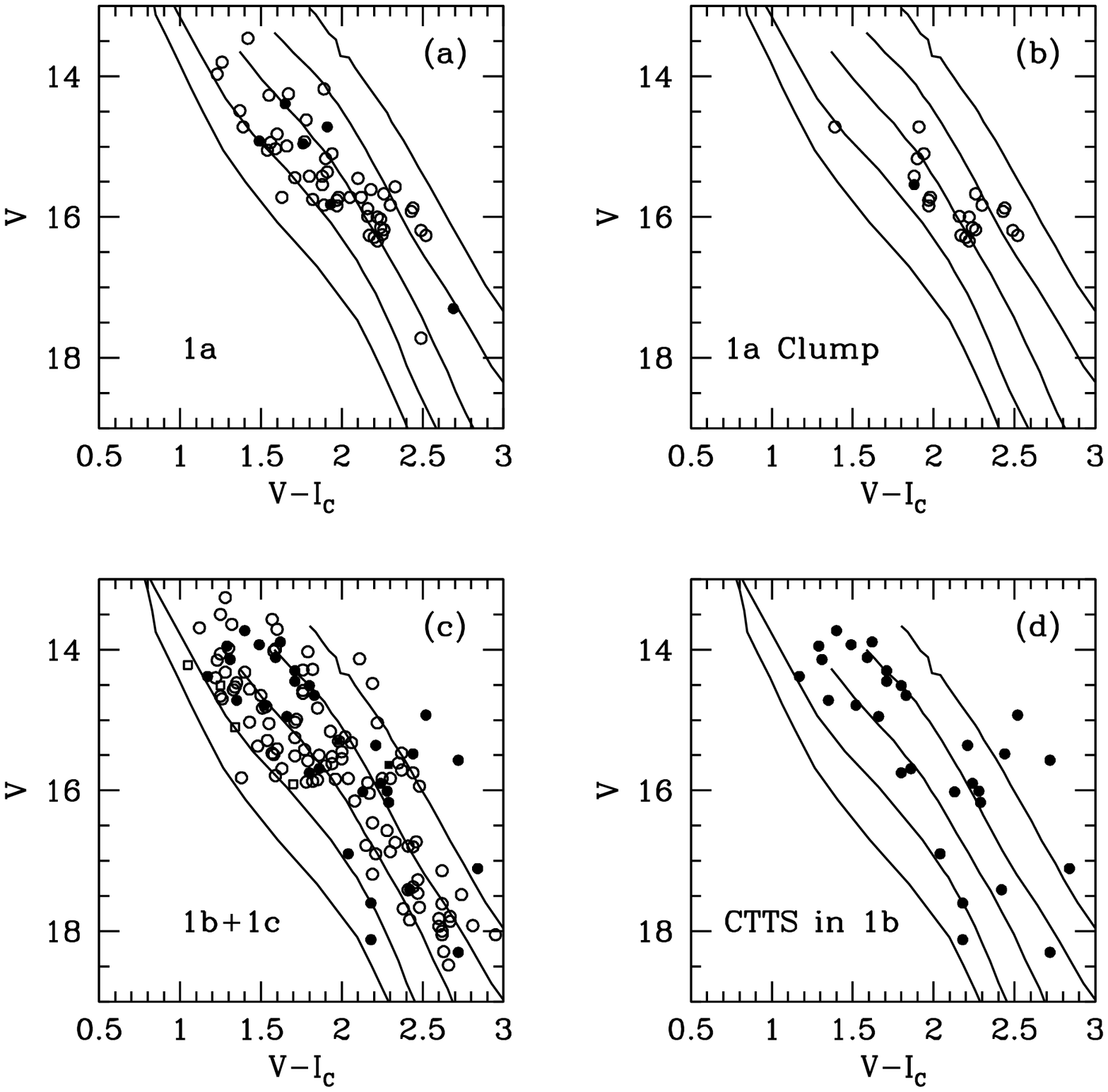}
\caption{Same as Figure \ref{cmds} using \citet{bca98}
isochrones.
}
\label{cmdsbarafe}
\end{figure}

\clearpage
\begin{figure}
\plotone{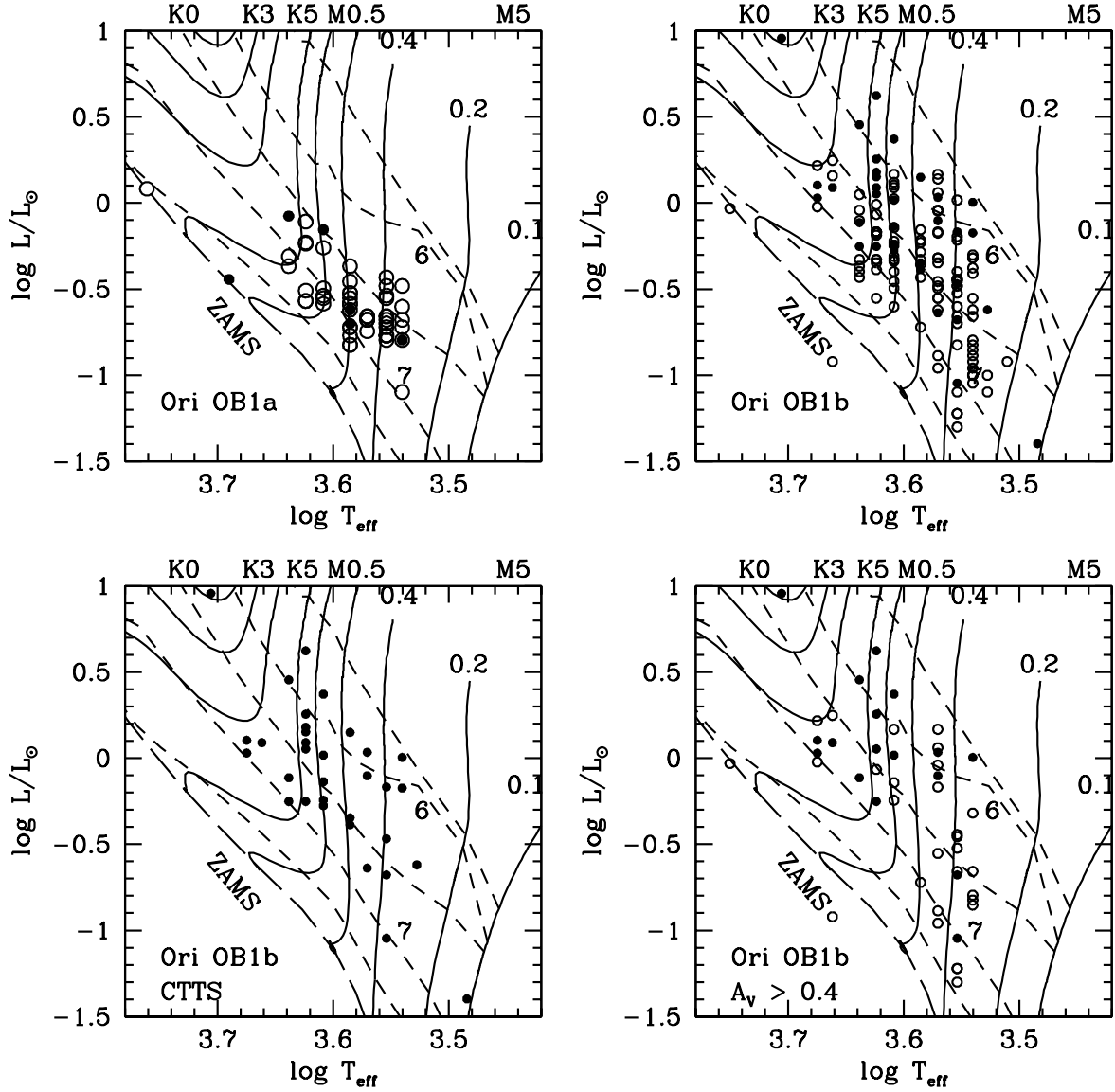}
\caption{
Location of the CTTS (solid circles)
and WTTS (open circles) in the H-R
 diagram.
Also shown are the Siess et al. (2000) evolutionary tracks (solid lines)
for masses
0.1$\msun$,
0.2$\msun$,
0.4$\msun$,
0.6$\msun$,
0.8$\msun$,
1.0$\msun$, and
1.5$\msun$, from left to right, and
and isochrones f (dotted lines) or
log age=5.5, 6, 6.5, 7, 7.5, from top
to bottom. The lower isochrone is the ZAMS.
Upper left: stars
in Ori OB1a; upper right: stars in Ori OB1b;
lower left: Ori OB1b CTTS; lower right: CTTS and WTTS
in Ori Ob1b with $A_V > 0.4$.
}
\label{hrsiess}
\end{figure}

\clearpage
\begin{figure}
\plotone{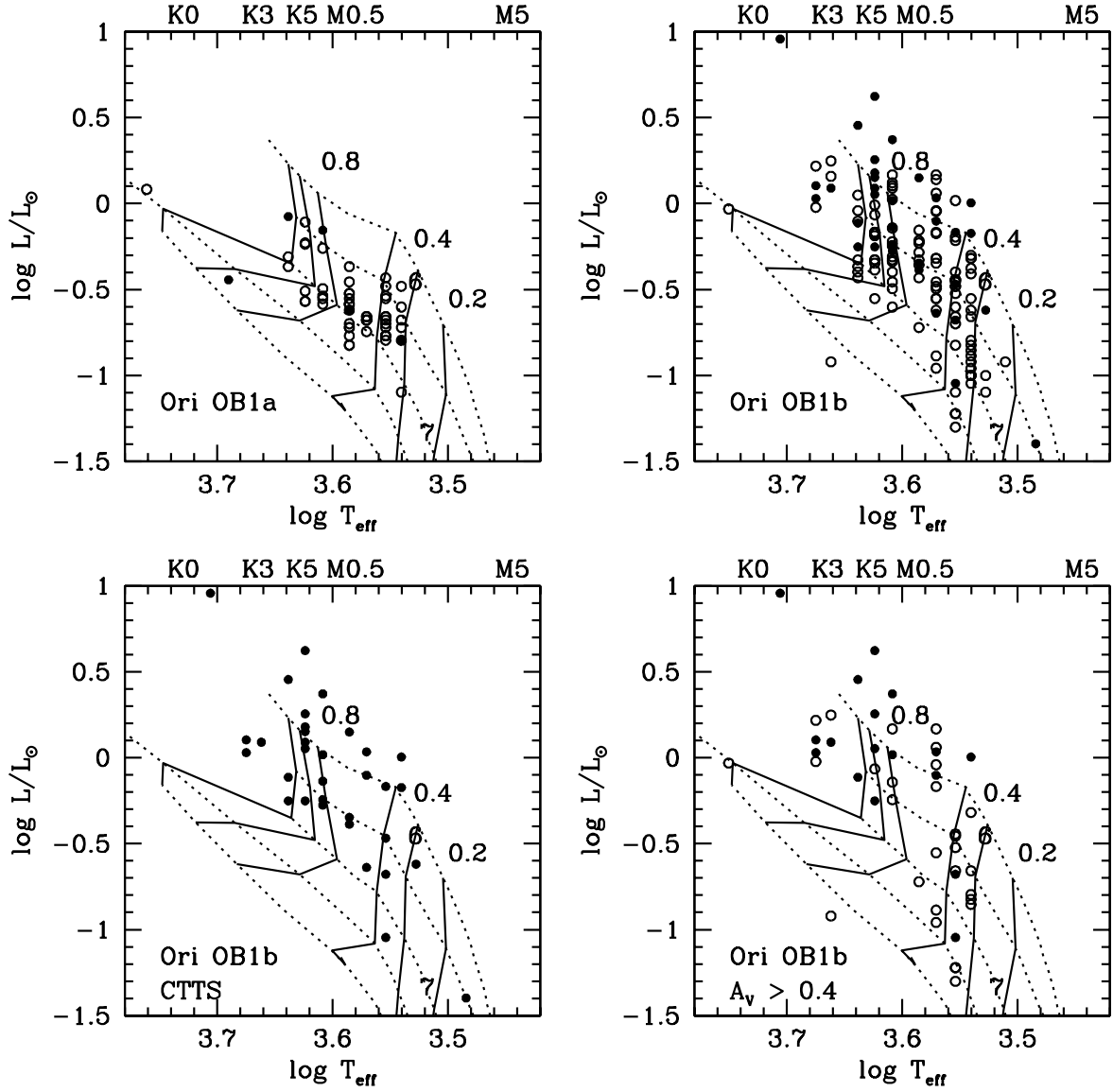}
\caption{
Similar to Figure \ref{hrsiess} using
Baraffe et al. (1998) evolutionary tracks and isochrones.
From left to right, evolutionary tracks (solid lines)
for masses
0.2$\msun$,
0.4$\msun$,
0.6$\msun$,
0.8$\msun$,
0.9$\msun$, and
1.0$\msun$.
From top to bottom
isochrones  (dotted lines) for
log age= 6, 6.5, 7, 7.5, and 8.
}
\label{hrbarafe}
\end{figure}

\clearpage
\begin{figure}
\plotone{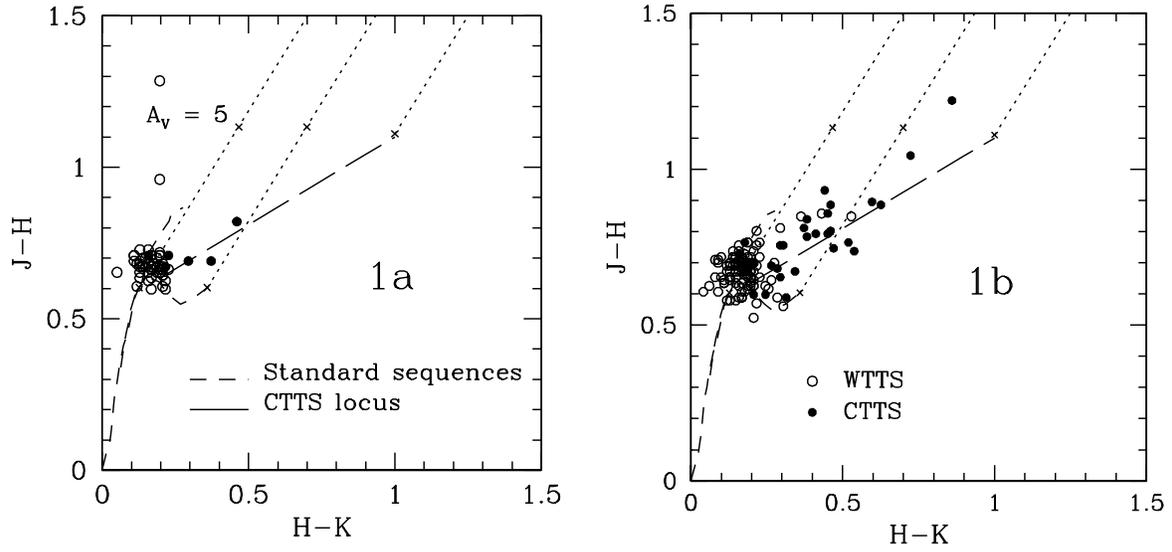}
\caption{
Location of CTTS (solid cicles)
and WTTS (open circles) in Ori OB1a (left)  and Ori OB1b (right) in the JHK diagram.
Colors are taken from 2MASS.
The standard main sequence and giant relations and reddening vector (for
$R_V$ = 3.1) with tickmarks
at intervals of $A_V$ = 5 are indicated, as well as
the CTTS locus (Meyer, Calvet, \& Hillenbrand 1997).
}
\label{jhk}
\end{figure}

\clearpage
\begin{figure}
\plotone{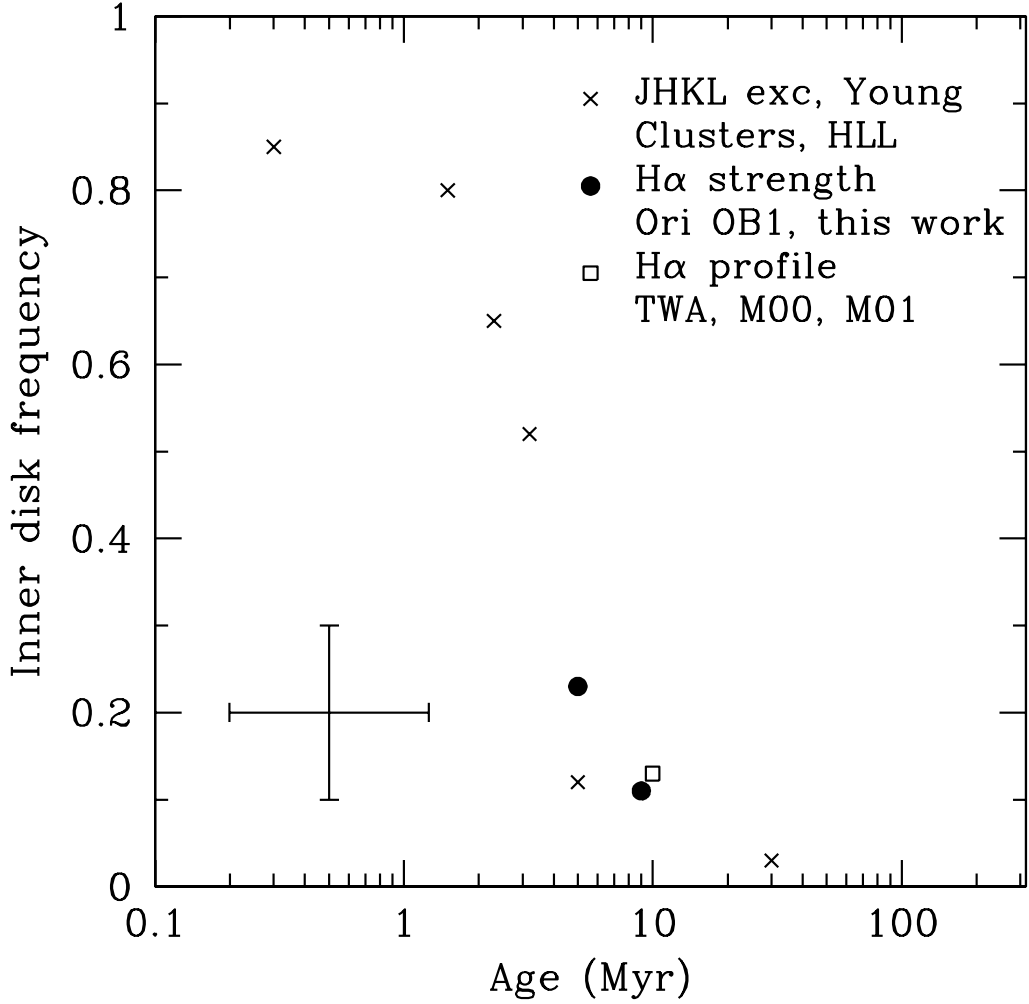}
\caption{Inner disk frequency as a function of age. The disk
frequencies have been estimated by different methods:
JHKL excesses in young clusters (Haisch, Lada, and Lada 2001, HLL);
H$\alpha$ strength, that is, relative number of CTTS
in Ori OB1a and 1b, in this work; broad H$\alpha$ profile
in TWA members (Muzerolle et al. 2000, 2001, M00, M01).
Typical errors are shown.
}
\label{diskfr}
\end{figure}


\begin{thebibliography}{}
 
\bibitem[Adams \& Myers(2001)]{adm01} Adams, F.C. \& Myers, P.C. 2001, \apj, 553, 744

\bibitem[Akerlof et al.(2000)]{akerlof00} Akerlof, C. et al. 2000, \aj, 119, 2001

\bibitem[Alcal\'a et al. (1996)]{atw96} Alcal\'a, J.M., Terranegra, L., Wichmann, R., et al. 1995, \aaps, 119, 7

\bibitem[Ali \& Depoy(1995)]{ali95} Ali, B.~\& Depoy, D.~L.\ 1995, \aj, 109, 709

\bibitem[Alcal\'a et al.(1996)]{atw96}  Alcal\'a, J.M., Terranegra, L., Wichmann. R., Chavarria-K., C., Krautter, J., Schmitt, J.H., Moreno, M., De lara, E. \& Wagner, R. 1996, \aap, 119, 7

\bibitem[Baltay et al.(2002)]{bsa02} Baltay, C., Snyder, J.A., Andrews, P. et al. 2002, \pasp, 114, 780

\bibitem[Baraffe et al.(1998)]{bca98} Baraffe, I., Chabrier, G., Allard, F., Hauschildt, P.H. 1998, \aap, 337, 403

\bibitem[Bally et al. (1987)]{bally87} Bally, 
J., Stark, A.~A., Wilson, R.~W., \& Langer, W.~D.\ 1987, \apjl, 312, L45 

\bibitem[Blaauw(1964)]{bla64}, Blaauw, A. 1964, \araa, 2, 213

\bibitem[Blaauw(1991)]{bla91} Blaauw, A. 1991, in The Physics of Star Formation and Early
Stellar Evolution, eds. C. Lada and N.D. Kylafis,
(Dordrecht: Kluwer), p. 125

\bibitem[Brice\~no et al. (1998)]{bhs98} Brice\~no,C ., Hartmann, L., Stauffer, J., Mart{\'\i}n, E. 198, \aj, 115, 2074

\bibitem[Brice\~no et al.(1997)]{bhs97} Brice\~no, C., Hartmann, L., Stauffer, J., Gagne, M., Caillault, J.-P., \& Stern, A. 1997, \aj, 113, 740

\bibitem[Brice\~no et al.(2001)]{bvc01} Brice\~no, C., Vivas, A.K., Calvet, N., Hartmann, L. et al. 2001, Science, 291, 93

\bibitem[Brown et al.(1994)]{bgz94}  Brown, A.G.A.,  de Geus, E.J., \& de Zeeuw, P.T. 1994, AA, 289, 101

\bibitem[Caldwell \& Falco(1993)]{caf93} Caldwell, N. \& Falco, E., 1993, Electronic 4Shooter/1.2m Manual

\bibitem[Cardelli, Clayton, \& Mathis(1989)]{ccm89} Cardelli, J.A., Clayton, G.C., Mathis, J.S. 1989, \apj, 345, 245

\bibitem[Dahari \& Lada(1999)]{dal99} Dahari, D. B., Lada, E. 1999,  \baas, 195, 7913

\bibitem[Dolan \& Mathieu(1999)]{dom99} Dolan, C. J., \& Mathieu, R., D. 1999, \aj, 118, 2409

\bibitem[Dolan \& Mathieu(2001)]{dom01} --- 2001, \aj, 121, 2124
 
\bibitem[Dolan \& Mathieu(2002)]{dom02} --- 2002, \aj, 123, 387

\bibitem[Fabricant et al.(1998)]{fhs98}  Fabricant, D.G., Hertz, E.N., Szentgyorgyi, A.H., Fata, R.G., Roll, J.B., \& Zajac, J.M. 1998,
Construction of the Hectospec: 300 optical fiber-fed spectrograph for the converted MMT,
Proc. SPIE, 3355, 285

\bibitem[Garmire et al. (2000)]{gfb00} Garmire, G., Feigelson, E.D., Broos, P. et al. 2000, \aj, 120, 1426

\bibitem[Garrison(1967)]{gar67} Garrison, R.F. 1967, \pasp, 79, 433

\bibitem[Genzel et al.(1981)]{grm81} Genzel, R., Reid, M. J., Moran, J.M., \& Downes, D. 1981, \apj, 244, 884
\bibitem[Genzel \& Stutzki(1989)]{ges89} Genzel, R. \& Stutzki, J. 1989, \araa, 27, 41

\bibitem[Gibson \& Hickson(1992)]{gih92}  Gibson, B.K. \& Hickson, P. 1992, \mnras, 258, 543

\bibitem[Haisch et al.(2001)]{hlp01}  Haisch, K.E., Jr., Lada, E.A., Pi\~na, R.K., Telesco, C.M., Lada, C.J.  2001, \aj, 121, 1512

\bibitem[Hartmann(1998)]{har98} Hartmann, L. 1998, Accretion Processes in Star Formation (Cambridge University Press)

\bibitem[Hern\'andez et al.(2003)]{hcb03} Hern\'{a}ndez J., Calvet N., Brice\~{n}o C., Hartmann L., and Berlind P., 2003, \aj, 127, 1682

\bibitem[Herbig(1962)]{her62} Herbig, G.H. 1962, Adv. Astron. Astrophys. 1, 47

\bibitem[Herbig \& Bell(1988)]{hbc88} Herbig, G.H., \& Bell, K.R. 1988, Lick Obs. Bull. 1111, HBC

\bibitem[Herbig \& Terndrup(1986)]{het86} Herbig, G.H. \& Terndrup, D.M. 1986, \apj, 307, 609

\bibitem[Herbst et al.(1994)]{hhg94} Herbst, W., Herbst, D.K., Grossman, E.J., \& Weinstein, e.j. 1994, \aj, 108, 1906

\bibitem[Hillenbrand(1997)]{hil97} Hillenbrand, L. 1997, \aj, 113, 1733

\bibitem[Joy(1945)]{joy45} Joy, A.H. 1945, \apj, 102, 168

\bibitem[Kenyon \& Hartmann(1995)]{kha95} Kenyon, S.  J., \& Hartmann, L., 1995, \apjs, 101,117

\bibitem[Kogure et al.(1989)]{kyw89} Kogure, T., Yoshida, S., Wiramihardja, S.,
 Nakano, M., Iwata, T. \& Ogura, K. 1989, PASJ, 41, 1195

\bibitem[Lada(1992)]{lad92} Lada, E. 1992, \apj, 393, 25

\bibitem[Lamm et al.(2004)]{lbm04} Lamm, M. H., Bailer-Jones, C. A. L., Mundt, R., Herbst, W., Scholz, A. 2004, \aap, 417, 557

\bibitem[Landolt(1992)]{landolt92}  Landolt, A.U. 1983, \aj, 88, 439

\bibitem[Lawson et al.(2001)]{lcm01} Lawson, W.A., Crause, L.A., Mamajek, E.E., Feigelson, E.D. 2001, \mnras, 321, 57

\bibitem[Maddalena et al.(1987)]{mmm87}  Maddalena, R. J., Morris, M., Moscowitz, J., \& Thaddeus, P. 1987, \apj, 303, 375

\bibitem[Mamajek et al.(2002)]{mml02} Mamajek, E.E., Meyer, M.R., \& Liebert, J., 2002, \aj, 124, 167

\bibitem[Meyer, Calvet, \& Hillenbrand(1997)]{mch97} Meyer, M.R., Calvet, N., Hillenbrand, L.A. 1997, \aj, 114, 288

\bibitem[Monet et al.(1998)]{mon98}  Monet, D. et al. 1998, in the PMM USNO-A2.0 Catalog (Washington: USNO)

\bibitem[Mora et al.(2001)]{2001A&A...378..116M} Mora, A., et al.\ 2001,
\aap, 378, 116

\bibitem[Motte et al.(2001)]{maw01} Motte, F., Andr\'e, P, Ward-Thompson, D., Bontemps, S.  2001, \aap, 372L, 41

\bibitem[Murdin \& Penston(1977)]{mup77} Murdin, P. \& Penston, M. 1977, \mnras, 181, 657

\bibitem[Muzerolle et al.(2001)]{2001ysne.conf..245M} Muzerolle, J., 
Hillenbrand, L., Calvet, N., Hartmann, L., \& Brice{\~ n}o, C.\ 2001, ASP 
Conf.~Ser.~244: Young Stars Near Earth: Progress and Prospects, 245 

\bibitem[Muzerolle et al.(2000)]{2000ApJ...535L..47M} Muzerolle, J., 
Calvet, N., Brice{\~ n}o, C., Hartmann, L., \& Hillenbrand, L.\ 2000, 
\apjl, 535, L47 

\bibitem[Palla \& Stahler(1992)]{pas92} Palla, F., \& Stahler, S.W. 1992, \apj, 392, 667

\bibitem[Palla \& Stahler(1993)]{pas93}  Palla, F., \& Stahler, S.W. 1993, \apj, 418, 414

\bibitem[Pojm\'anski(2003)]{pojmanski03} Pojm\'anski, G.  2003, AcA, 53, 341

\bibitem[Prosser et al.(1994)]{psh94} Prosser, C.F., Stauffer, J.R., Hartmann, L., Soderblom, D.R., Jones, B.F., Werner, M.W., McCaughrean, M.J. 1994, \apj, 421, 517

\bibitem[Preibisch et al.(2002)]{pbb02} Preibisch, T., Brown, A.G.A., Bridges, T. Guenther, E., Zinnecker, H. 2002, \aj, 124, 404

\bibitem[Preibisch \& Zinnecker(1999)]{prz99} Preibisch, T. \& Zinnecker, H. 1999, \aj, 117, 238

\bibitem[Preibisch et al.(1997)]{pzg97} Preibisch, T., Zinnecker, H., G\"unther, E., Sterzik, M. F., Frink, S., R\"oser, S., Kunkel, M, 1997, Astron. Ges., Abstr. Ser. 13, 18

\bibitem[Nakano et al.(1995)]{nwk95} Nakano, M., Wiramihardja, S. D., \& Kogure, T. 1995, PASJ, 49, 889

\bibitem[Rengstorf et al.(2004)]{rma04}  Rengstorf, A. et al. 2004, \apj, submitted.

\bibitem[Sabbey et al.(1998)]{sco98}  Sabbey, C.N., Coppi, P. \& Oemler, A. 1998, \pasp, 110, 1067

\bibitem[Sherry et al.(1999)]{sww99} Sherry., W., Walter, F. M., \& Wolk, S. J.  1999, \baas 194, 6824

\bibitem[Sherry(2003)]{she03} Sherry, W. 2003, Ph.D. Thesis.

\bibitem[Siess et al.(2000)]{sdf00}  Siess, L., Dufour, E., \& Forestini, M.\ 2000, \aap, 358, 593

\bibitem[Schlegel et al.(1998)]{sfdd98} Schlegel, D.J., Finkbeiner, D.P., Davis, M. 1998, \apj, 500, 525

\bibitem[Snyder(1998)]{sny98} Snyder, J.A. 1998, SPIE, 3355, 635

\bibitem[Stassun et al.(1999)]{smm99} Stassun, K.G., Mathieu, R.D., Mazeh, T., Vrba, F.J. 1999, \aj, 117, 2941

\bibitem[Stock \& Abad(1988)]{stockabad88}  Stock, J. \& Abad, C. Publ. 1988, CIDA TH-161

\bibitem[van den Ancker, de Winter, \& Tjin A
Djie(1998)]{1998A&A...330..145V} van den Ancker, M.~E., de Winter, D., \&
Tjin A Djie, H.~R.~E.\ 1998, \aap, 330, 145


\bibitem[Vivas et al.(2004)]{vza04}  Vivas, A.K. et al. 2004, \aj, 127, 1158
 
\bibitem[Walter et al.(1997)]{wwf97}  Walter, F. M., Wolk, S. J., Freyberg, M., Schmitt, J. H. M. 1997, MmSAI, 68, 1081

\bibitem[Walter, Wolk \& Sherry(1998)]{wws98} Walter, F. M., Wolk, S. J., \& Sherry, W. 1998, ASP Conf. Ser. 154, eds. R. Donahue and J. Bookbinder, p. 1793
 
\bibitem[Walter et al.(2000)]{wan00} Walter, F.M., Alcal\'a, J.M, Neuh\"auser, R. Sterzik, M. \& Wolk, S. 2000, in Protostars and Planets III, eds.
E.H. Levy \& J.I. Lunine (Tucson: University of Arizona Press), p. 273

\bibitem[Warren \& Hesser(1977)]{wah77} Warren, W.H., \& Hesser, J.E. 1977, \apjs, 34, 115

\bibitem[Warren \& Hesser(1978)]{wah78} Warren, W.H., \& Hesser, J.E. 1978, \apjs, 36, 497

\bibitem[White \& Basri(2003)]{wba03} White, R.J. \& Basri, G. 2003, \apj, 582, 1109

\bibitem[Wiramihardja et al.(1989)]{wky89} Wiramihardja, S., Kogure, T., Yoshida, S., Ogura, K.,
\& Nakano, M. 1989, in Interstellar Dust, NASA N91-14897 06-88, p. 239

\bibitem[Wiramihardja et al.(1991)]{wky91} Wiramihardja, S., Kogure, T., Yoshida, S., Nakano, M., Ogura, K., \& Iwata, T. 1991, PASJ, 43, 27

\bibitem[Wiramihardja et al.(1993)]{wky93} Wiramihardja, S., Kogure, T., Yoshida, S., Ogura, K., \& Nakano, M. 1993, PASJ, 45, 643

 \end{thebibliography}
\end{document}